\documentclass[titlepage,12pt,subeqn]{utarticle}
\usepackage{amsmath,amssymb,amscd}
\usepackage{color}
\usepackage{latexsym}
\usepackage{ifthen}
\usepackage[xdvi]{graphics}
\usepackage[]{hyperref} 
\usepackage{bbm}

\numberwithin{equation}{section}

\newcommand{\sign}{\operatorname{sign}}

\newcommand{\ic}{{\rm i}}

\def\eps{\epsilon}

\begin{document}
\preprint{
UTTG-05-06 \\
{\tt hep-th/0604087}\\
}
\title{Towards a Realistic Type IIA $T^6/{\mathbb Z}_4$
Orientifold Model with Background Fluxes, Part 1: Moduli Stabilization}

\author{ {\sc Matthias Ihl} and {\sc Timm Wrase}
     \oneaddress{
      Theory Group, Department of Physics,\\
      University of Texas at Austin,\\
      Austin, TX 78712, USA \\
      {~}\\
      \email{matze, wrasetm@physics.utexas.edu}\\
      }
}

\date{\today}

\Abstract{We apply the methods of DeWolfe {\it et al.}
[hep-th/0505160] to a $T^6/{\mathbb Z}_4$ orientifold model. This is
the first step in an attempt to build a phenomenologically
interesting meta-stable de Sitter model with small cosmological
constant and standard model gauge groups. }

\maketitle

\section{Introduction}
There have been some longstanding, unsolved problems when it comes to
realistic model building within the framework of Calabi-Yau (CY)
compactifications of superstring theories, namely
\begin{itemize}
\item supersymmetry breaking,
\item the moduli problem,
\item a small, but nonvanishing cosmological constant $\Lambda > 0$
\cite{Seljak:2004xh,Riess:2004nr,Spergel:2003cb},
indicating an asymptotic de Sitter (dS) type universe. Moreover,
$w < -\frac{1}{3}$ indicates an accelerated expansion.
\end{itemize}
Especially the last point seems to pose a serious challenge for
string theory, because (eternal) de Sitter type universes, due to
the existence of event horizons, are believed to necessitate a
finite number of physical degrees of freedom (resulting in finite
dimensional Hilbert spaces) \cite{Fischler:2001yj,Hellerman:2001yi},
which appears impossible to reconcile with string theory. Moreover,
compactifications of string theory on Calabi-Yau 3-folds to four
spacetime dimensions generically produce a large number of massless
moduli (scalar fields) which we do not observe in nature. However, a
recent proposal by Kachru, Kallosh, Linde and Trivedi (KKLT)
\cite{Kachru:2003aw} manages to address all of the above stated
difficulties at once. The authors outline a way to produce a
nontrivial (scalar) potential for {\it all} CY moduli, resulting in
supersymmetric anti-de Sitter (AdS) vacua in which all moduli are
stabilized. To achieve this, the authors start with a warped
compactification of a type IIB orientifold with background fluxes as
discussed in \cite{Giddings:2001yu}. There it was shown that by
turning on appropriate R-R and NS-NS 3-form fluxes $\hat{F}_{(3)}$
and $\hat{H}_{(3)}$, it is possible to fix both the complex
structure moduli $z^{\alpha}$ and the axiodilaton $\tau :=
C_{(0)}+\ic e^{-\hat{\phi}}$. However, owing to the fact that the
flux-induced superpotential\footnote{Here we introduce the
complexified 3-flux $\hat{G}_{(3)}:= \hat{F}_{(3)}-\tau
\hat{H}_{(3)}$.} $W_0^{IIB}=\int_{CY_3} \hat{G}_{(3)} \wedge \Omega$
\cite{Gukov:1999ya} does not depend on the K{\"a}hler moduli of the
compactification manifold, one is forced to include nonperturbative
corrections to $W$ in order to generate a potential for those
moduli. KKLT argue that this can be achieved generically in their
class of models by one of two effects: Euclidean $D3$-brane
instantons wrapping divisors of arithmetic genus equal to one
\cite{Witten:1996bn} or gaugino condensation in the gauge theory
living on a stack of coinciding $D7$-branes wrapping 4-cycles of the
internal CY \cite{Veneziano:1982ah,Taylor:1982bp}. Both effects can
be shown to lead to stabilization of the remaining K{\"a}hler
moduli. As a matter of fact, the condition on the arithmetic genus
of the divisors can be relaxed in the presence of fluxes, as was
discovered recently by several authors (see
e.g.~\cite{Berglund:2005dm}). In the final step of the KKLT
construction it is argued that by adding $\overline{D3}$-branes to
the setup in a suitable fashion, it is possible to break
supersymmetry in such a way that the vacuum is lifted to a dS vacuum
with a discretely tunable cosmological constant\footnote{This tuning
can be achieved by turning on appropriate fluxes through cycles in
the internal manifold.}. It is, however, important to note that the
dS vacua in question are only local minima of the ${\cal N}=1$
supergravity scalar potential for the relevant moduli. There always
exists a global minimum, the Dine-Seiberg runaway vacuum in the
large volume or decompactification limit. Therefore the dS vacua are
only metastable, albeit at cosmological time scales, thus evading
the above mentioned problems concerning eternal de Sitter
spacetimes. \\
The program outlined by KKLT triggered a myriad of work within the
framework of type IIB orientifold compactifications
\cite{Denef:2005mm,Gorlich:2004qm,Lust:2005dy,Reffert:2005mn}.
Several important refinements to the original KKLT proposal were
made, e.g., V. Balasubramanian, F. Quevedo and collaborators
\cite{Balasubramanian:2004uy, Balasubramanian:2005zx,Conlon:2005ki}
realized that it is inconsistent (at least generically) to neglect
the perturbative $\alpha'$-corrections to the K{\"a}hler potential.
Stated differently, by including these corrections, one can prove
the existence of AdS vacua (even nonsupersymmetric ones) and the
validity of the construction for a much broader range of parameters
as compared to the original proposal without perturbative
corrections.\\
In recent months, several authors have studied various aspects of
the KKLT program in the framework of type IIA orientifold
compactifications
\cite{Villadoro:2005cu,DeWolfe:2005uu,Kachru:2004jr,Camara:2005dc,Saueressig:2005es}.
One important difference compared to the type IIB case is that here,
as we shall see below, the flux-induced superpotential $W_0^{IIA}$
contains contributions both from the complex structure as well as
the K{\"a}hler moduli. Therefore it is possible to stabilize both
types of moduli\footnote{One can stabilize all the complex structure
moduli but only one linear combination of the axions.} without
having to consider nonperturbative instanton corrections. Another
worthwhile observation is that whereas in the type IIB scenario the
fluxes are highly constrained by the tadpole cancelation condition
for the $\hat{C}_{(4)}$-field, this is not true in the IIA setup,
where some of the fluxes, namely $\hat{F}_{(2)}$ and
$\hat{F}_{(4)}$, are left unaffected and thus unconstrained by the
$\hat{C}_{(7)}$-tadpole cancelation condition
\cite{Villadoro:2005cu,DeWolfe:2005uu,Camara:2005dc}.\\
In the present paper we work out and discuss in some detail the
moduli stabilization for a specific $T^6/{\mathbb Z}_4$ orientifold
model. It has the prospect of yielding a viable stringy realization
of the ingredients needed for a realistic description of particle
physics, namely the correct particle spectrum (SM or MSSM) combined
with desired cosmological features ($\Lambda >0$). These more
advanced issues will be addressed in future research. In this paper
we find supersymmetric and nonsupersymmetric AdS vacua in which all
moduli are stabilized. Moreover we exhibit some vacua in which one
K{\"a}hler modulus remains unfixed (flat direction), although we
have turned on generic fluxes.\\
The paper is organized as follows: We begin by introducing the basic
setup and the construction of the orientifold model in
section~\ref{sec:basic}. Section~\ref{sec:moduli} contains a
detailed discussion of moduli stabilization via flux-induced
potentials for the moduli of the untwisted sector. We present two
different approaches to this problem: First, starting from
ten-dimensional massive type IIA supergravity, we obtain the
four-dimensional effective scalar potential by Kaluza-Klein
reduction. Second, we solve supersymmetric F-flatness conditions in
the language of four-dimensional ${\mathcal N}=1$ supergravity,
yielding supersymmetric AdS vacua. We extend our considerations to
the twisted sector moduli fields in section~\ref{sec:twisted},
followed by some conclusions and an outlook in
section~\ref{sec:conclusions}.

\section{Basic setup}\label{sec:basic}

\subsection{The $T^6/{\mathbb Z}_4$ orientifold}

In this section, we outline the properties of the type IIA
orientifold model under investigation, namely an orientifolded
$T^6/{\mathbb Z}_4$ orbifold that preserves ${\cal N}=1$
supersymmetry. A detailled discussion of this model can be found in
\cite{Blumenhagen:1999ev}.\\
\noindent {\bf The $T^6/{\mathbb Z}_4$ orbifold.} As a first step,
we want to compactify type IIA string theory on an $T^6/{\mathbb
Z}_4$ orbifold background\footnote{The $T^6/{\mathbb Z}_4$ orbifold
is among those studied in \cite{Dixon:1985jw,Dixon:1986jc} and has
been shown to admit consistent string propagation, e.g., preserving
modular invariance.}. Let us start by describing the orbifold
construction, following
\cite{Blumenhagen:1999ev,Blumenhagen:2002gw}. It is important to use
a lattice for the $T^6$ that implements a crystallographic action of
the cyclic group. Therefore one chooses the root lattice of an
appropriate Lie algebra. In the ${\mathbbm Z}_4$ case under
investigation the appropriate choice is $SU(2)^6$. Unlike the more
complicated orbifolds with quotient group ${\mathbbm Z}_N$ for $N>6$
\cite{Blumenhagen:2004di}, in the case of ${\mathbbm Z}_4$, the root
lattice of the Lie algebra allows a choice of complex structure in
such a way that the torus factorizes as $T^6=T^2_{(1)} \times
T^2_{(2)} \times T^2_{(3)}$. We parameterize it by three complex
coordinates $z^i, \; i \in \{1,2,3\}$, together with the periodic
identifications
\begin{equation}
z^i \sim z^i + \pi_{2i-1} \sim z^i + \pi_{2i}, \; i \in \{1,2,3\},
\end{equation}
where the $\pi_k$ denote the fundamental 1-cycles of the three 2-tori.
The ${\mathbbm Z}_4$ action on the torus $T^6$ is given by
\begin{equation}
\Theta: (z^1,z^2,z^3) \mapsto (\alpha z^1, \alpha z^2, \alpha^{-2} z^3),
\end{equation}
where $\alpha= e^{\ic \pi/2}=\ic$ is a fourth root of unity and
$\Theta^4 = {\mathbbm 1}$. This action preserves ${\cal N}=2$
supersymmetry in four dimensions, implying that the orbifold is
actually a singular limit of a Calabi-Yau 3-fold. The Hodge numbers
are given by $h^{1,1}=31$ and $h^{2,1}=7$, yielding the number of
K{\"a}hler and complex structure moduli before the orientifold
projection. Table 1 lists how the complex structure and K{\"a}hler
moduli appear in the different sectors of the orbifold.

\vskip 0.8cm
\vbox{ \centerline{\vbox{ \hbox{\vbox{\offinterlineskip
\def\tablespace{height2pt&\omit&&\omit&&\omit
&&\omit&\cr}
\def\tablerule{\tablespace\noalign{\hrule}\tablespace}
\def\tableruleA{\tablespace\noalign{\hrule height1pt}\tablespace}
\hrule\halign{&\vrule#&\strut\hskip0.2cm\hfill #\hfill\hskip0.2cm\cr
&sector:  && untwisted && $\Theta,\Theta^3$-twisted && $\Theta^2$-twisted
&& $\sum$ &\cr
\tablerule
& fixed points/type: && --- && 16 ${\mathbb Z}_4$ && 12 ${\mathbb Z}_2$+
4 ${\mathbb Z}_4$ (${\mathbb Z}_2$) && --- &\cr
\tablerule
& complex structure:
     && 1 && --- && 6+0 && 1+6 &\cr
\tablerule & K{\"a}hler:  && 5 && 16 && 6+4 && 5+26 &\cr }\hrule}}}}
\centerline{ \hbox{{\bf Table 1:}{\it ~~ List of complex structure
and K{\"a}hler moduli. }}} } \vskip 0.5cm \noindent The Euler
characteristic turns out to be
\begin{equation}
\chi (T^6/{\mathbbm Z}_4)= 2(h^{1,1}-h^{2,1})=
\frac{1}{|{\mathbbm Z}_4|}\sum_{gh=hg}\chi(g,h)
= 48,
\end{equation}
where $\chi(g,h)$ denotes the Euler characteristic of the subspace invariant
under both $g$ and $h$. $|{\mathbbm Z}_4|=4$ is the order of the group. The
sum runs over all pairs of elements of the
Abelian subgroup of the quotient group; here, since ${\mathbbm Z}_4$ is
Abelian, the sum runs over the sixteen pairings involving all four group
elements\footnote{The actions of $\Theta^1, \Theta^2, \Theta^3$ all yield
16 fixed points. However, four pairs of elements, namely those involving
combinations of $\Theta^0={\mathbbm 1}$ and $\Theta^2: (z^1,z^2,z^3)
\mapsto (\alpha^2 z^1, \alpha^2 z^2, z^3)$, leave at least one of the $T^2$
factors invariant, thus not contributing to the sum, as $\chi(T^6)=
\chi(T^2)=0$.}.\\
\noindent {\bf The orientifold model.} As in
\cite{Blumenhagen:1999ev,Blumenhagen:2002gw}, we construct a
$T^6/{\mathbbm Z}_4$ orientifold by modding out by ${\mathcal
O}=\Omega_p (-1)^{F_L} \sigma$, where $\Omega_p$ denotes worldsheet
parity and $(-1)^{F_L}$ stands for left-moving fermion number. There
are two distinct choices for the antiholomorphic\footnote{In type
IIA superstring theory, the involutive symmetry $\sigma$ has to be
chosen to be antiholomorphic, since the left-moving space-time
supercharge corresponds to the holomorphic 3-form, whereas the
right-moving space-time supercharge corresponds to the
antiholomorphic 3-form. In the type IIB case both supercharges are
related to the holomorphic 3-form, thus necessitating a holomorphic
involution.\cite{Acharya:2002ag}} involution $\sigma$ on each of the
$T^2$. We choose\footnote{This is the {\bf ABB} model discussed in
detail in \cite{Blumenhagen:2002gw}.}
\begin{align}
&\sigma: z^1 \mapsto \bar{z}^1,\\ \nonumber &\sigma: z^2 \mapsto
\alpha \bar{z}^2, \\\nonumber  &\sigma: z^3 \mapsto \bar{z}^3.
\end{align}
For the first two tori, the complex structure is fixed to be $\ic$,
so $z^i= x^i +\ic y^i, \: i=1,2$. On the third torus the ${\mathbbm
Z}_4$ action does not fix the complex structure $z^3=x^3+ \ic U_2
y^3$. The tori and our choices of fundamental 1-cycles are shown in
figure 1.
\begin{figure}[h]
\begin{center}
\resizebox{10.6cm}{5.3cm}{\includegraphics{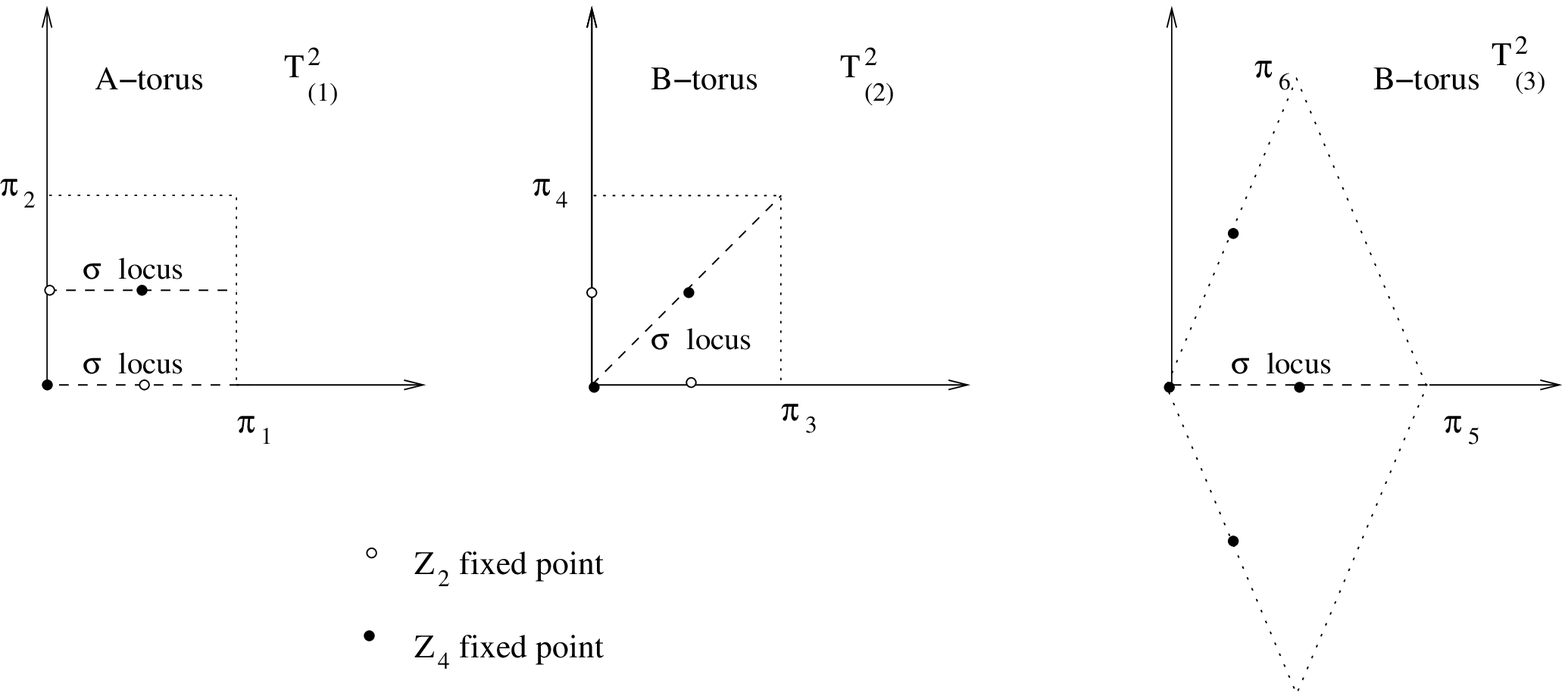}}
\setlength{\unitlength}{1cm}\\
{{\bf Figure 1:} Tori of the {\bf ABB} model.}
\end{center}
\end{figure}
After the orientifold projection we have an $O6$ orientifold plane
wrapping the invariant special Lagrangian 3-cycles in $T^6/{\mathbb
Z}_4$ and filling the four noncompact dimensions. For reference, we
have summarized the invariant cycles in each sector in table 2.
There we employ the notation
\begin{equation}
\pi_{ijk}:= \pi_i \otimes \pi_j \otimes \pi_k,
\end{equation}
where $\pi_{2i-1}$ and $\pi_{2i}$ denote the two fundamental
1-cycles of the three 2-tori $T^2_i, \; i \in \{1,2,3\}$ (see
figure 1).
\vskip 0.8cm \vbox{ \centerline{\vbox{
\hbox{\vbox{\offinterlineskip
\def\tablespace{height2pt&\omit&&\omit &\cr}
\def\tablerule{\tablespace\noalign{\hrule}\tablespace}
\def\tableruleA{\tablespace\noalign{\hrule height1pt}\tablespace}
\hrule\halign{&\vrule#&\strut\hskip0.2cm\hfill #\hfill\hskip0.2cm\cr
&projection  && fixed point set  &\cr
\tablerule
& ${\mathcal O}$ && $2 (\pi_{135}+\pi_{145})$ &\cr
\tablerule
& ${\mathcal O}\Theta$
     &&  $ 2 \pi_{145}+2 \pi_{245}-4 \pi_{146}-4\pi_{246}$ &\cr
\tablerule
& ${\mathcal O}\Theta^2$ && $2(\pi_{235}-\pi_{245})$ &\cr
\tablerule
& ${\mathcal O}\Theta^3$
     &&  $ -2 \pi_{135}+2 \pi_{235}+4 \pi_{136}-4\pi_{236}$ &\cr
}\hrule}}}} \centerline{ \hbox{{\bf Table 2:} Invariant cycles in
each sector of the {\bf ABB} model. }} } \vskip 0.5cm \noindent The
${\mathbbm Z}_4$ action maps the cycles invariant under ${\mathcal
O}$ and ${\mathcal O}\Theta^2$ into each other and likewise for the
other two cycles. Therefore there are two invariant 3-cycles that
are both wrapped once by the $O6$-plane:
\begin{align}
& [a_0] := 2 (\pi_{135}+\pi_{145}+\pi_{235}-\pi_{245})\\
& [a_1] := 4 (\pi_{136}- \pi_{146}- \pi_{246} -\pi_{236}) + 2(-
\pi_{135} + \pi_{145}+ \pi_{245} + \pi_{235})
\end{align}
In addition, there will be exceptional 3-cycles related to the
blow-ups of the fixed point singularities (cf.~section
\ref{sec:twisted}).\\
The $O6$-plane contributes to a $\hat{C}_{(7)}$-tadpole that has to
be canceled either by introducing $D6$-branes or by turning on
appropriate fluxes. This issue will be addressed  in the next
section. It is important to note that both the $O6$-plane and the
$D6$-branes can be chosen to preserve/break the same supersymmetry.
Thus, we are left with ${\cal N}=1$ supersymmetry in four
dimensions.

\subsection{Moduli and fluxes}

Before embarking on the task of generating appropriate potentials by
turning on fluxes, let us collect the relevant moduli fields, forms
and cycles appearing in our construction. We start out by taking a
closer look at the 3-cycles in the game. Since $b_{\mathrm{untw.}}^3
= 2+2 h_{\mathrm{untw.}}^{2,1}=4$, we expect four 3-cycles from the
untwisted sector. This fits nicely with the observation that the
only $(2,1)$-form invariant under the ${\mathbbm Z}_4$-action is
$dz^1\wedge dz^2 \wedge d \overline{z}^3$, so that the four 3-cycles
are simply the duals of the holomorphic $(3,0)$-form $\Omega$, the
antiholomorphic $(0,3)$-form $\overline{\Omega}$, the ${\mathbbm
Z}_4$-invariant $(2,1)$-form and the associated
${\mathbbm Z}_4$-invariant $(1,2)$-form. \\
The 1-cycles yield the following behavior under the ${\mathbbm
Z}_4$-action,
\begin{align}
&\Theta^1:& \pi_1 \mapsto +\pi_2, \pi_3 \mapsto +\pi_4, \pi_5
\mapsto -\pi_5,\\\nonumber &         & \pi_2 \mapsto -\pi_1, \pi_4
\mapsto -\pi_3, \pi_6 \mapsto -\pi_6,\\\nonumber &\Theta^2:& \pi_1
\mapsto -\pi_1, \pi_3 \mapsto -\pi_3, \pi_5 \mapsto
+\pi_5,\\\nonumber &         & \pi_2 \mapsto -\pi_2, \pi_4 \mapsto
-\pi_4, \pi_6 \mapsto +\pi_6,\\\nonumber &\Theta^3:& \pi_1 \mapsto
-\pi_2, \pi_3 \mapsto -\pi_4, \pi_5 \mapsto -\pi_5,\\\nonumber &
& \pi_2 \mapsto +\pi_1, \pi_4 \mapsto +\pi_3, \pi_6 \mapsto -\pi_6,
\end{align}
leading to the following ${\mathbbm Z}_4$-invariant combination of 3-cycles
\begin{align}
\rho_1&:=2(\pi_{135}-\pi_{245}),&\widetilde{\rho}_1:=
2(\pi_{136}-\pi_{246}),\\\nonumber
\rho_2&:=2(\pi_{145}+\pi_{235}),&\widetilde{\rho}_2:=
2(\pi_{146}+\pi_{236}).
\end{align}
Recall from table 1 that before the orientifold projection there are
in addition 5 K{\"a}hler moduli from the untwisted sector. \\
Next, we need to take a closer look at the moduli coming from the
twisted sectors. The $\Theta^1$- and the $\Theta^3$-twisted sectors
feature 16 ${\mathbbm Z}_4$ fixed points, giving rise to 16
additional K{\"a}hler moduli. The $\Theta^2$ action leaves the third
torus invariant, but acts nontrivially on the first two. Of the
sixteen ${\mathbbm Z}_2$ fixed points there are four that are also
fixed points under the ${\mathbbm Z}_4$-action. To each of the
sixteen fixed points we associate an exceptional 2-cycle $e_{\alpha
\beta}, \; \alpha , \beta \in \{1,2,3,4\}$, where $\alpha = 1,4$
denote the ${\mathbbm Z}_4$-invariant fixed points and $\alpha =
2,3$ denote the ${\mathbbm Z}_2$-invariant fixed points that get
mapped into each other under $\Theta$ (cf. figure 2). These give a
total of 10 K{\"a}hler moduli.
\begin{figure}[h]
\begin{center}
\resizebox{10.6cm}{5.3cm}{\includegraphics{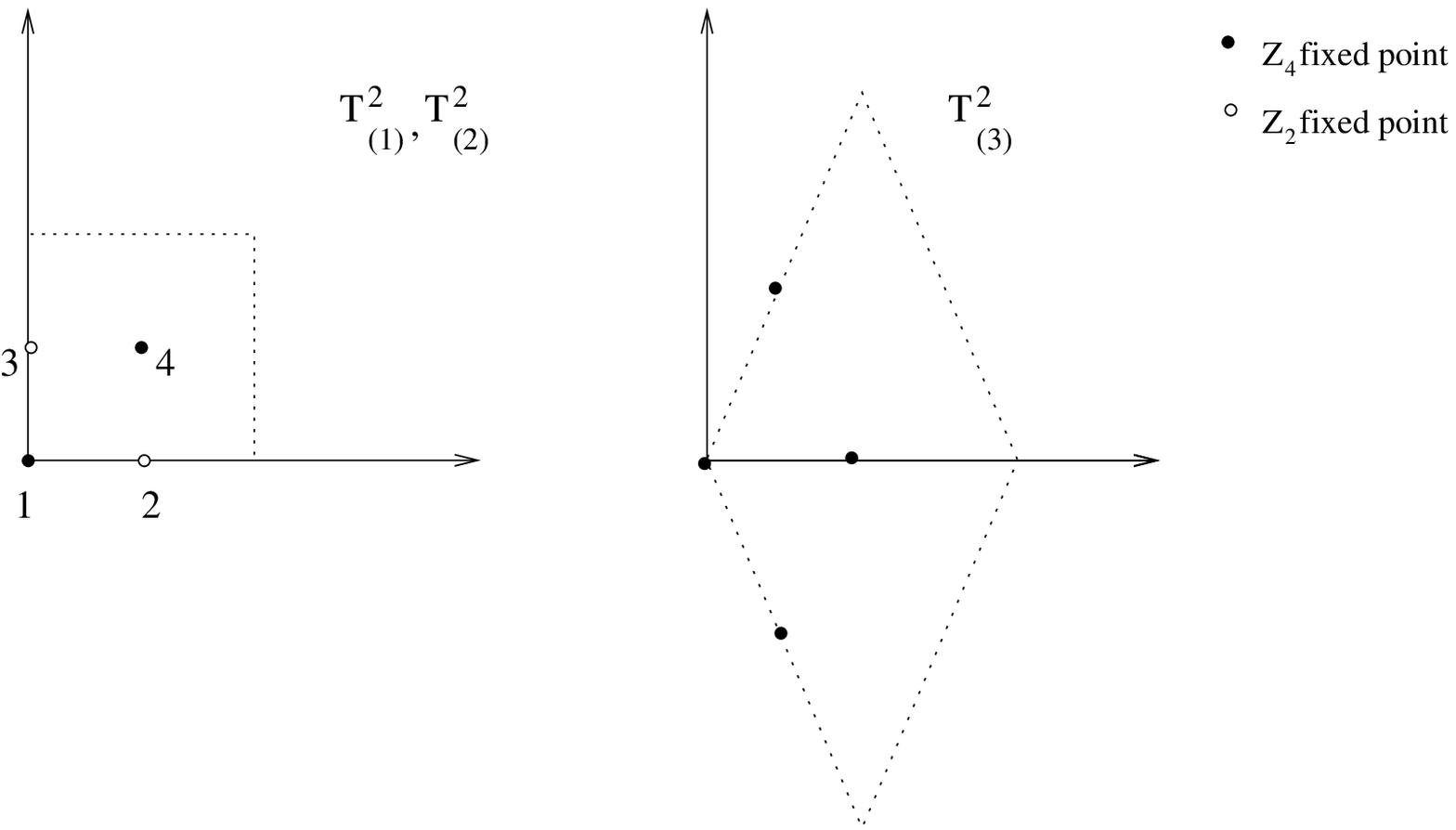}}
\setlength{\unitlength}{1cm}\\
{\bf Figure 2:} Fixed points of the first two tori and the third torus.
\end{center}
\end{figure}
Certain linear combinations of these 2-cycles may be combined with
the fundamental 1-cycles $\pi_{5,6}$ on the third torus to yield
exceptional 3-cycles of topology $S^2 \times S^1$. Demanding
invariance of the exceptional 3-cycles under the action of $\Theta$
and $\Theta^3$, which is given by \footnote{The two ${\mathbbm Z}_2$
fixed points are interchanged under $\Theta$ and $\Theta^3$, while
the ${\mathbbm Z}_4$ fixed points are invariant (cf.~figure 2). The
minus sign in (\ref{eq:excep3cycles}) stems from the reflection of
the fundamental 1-cycle of the third torus.}
\begin{align}\label{eq:excep3cycles}
\Theta(e_{\alpha \beta}\otimes \pi_{5,6})
=\Theta^3(e_{\alpha \beta}\otimes \pi_{5,6})= - e_{\eta(\alpha)\eta(\beta)}
\otimes \pi_{5,6},
\end{align}
with
\begin{equation}
\eta(1)=1,\; \eta(4)=4,\; \eta(2)=3,\; \eta(3)=2,
\end{equation}
one finds precisely twelve invariant combinations,
\begin{align}
&\eps_1 := (e_{12}-e_{13})\otimes \pi_5, & \widetilde{\eps}_1 :=
(e_{12}-e_{13})\otimes \pi_6,\\\nonumber &\eps_2 :=
(e_{42}-e_{43})\otimes \pi_5, & \widetilde{\eps}_2 :=
(e_{42}-e_{43})\otimes \pi_6,\\\nonumber &\eps_3 :=
(e_{21}-e_{31})\otimes \pi_5, & \widetilde{\eps}_3 :=
(e_{21}-e_{31})\otimes \pi_6,\\\nonumber &\eps_4 :=
(e_{24}-e_{34})\otimes \pi_5, & \widetilde{\eps}_4 :=
(e_{24}-e_{34})\otimes \pi_6,\\\nonumber &\eps_5 :=
(e_{22}-e_{33})\otimes \pi_5, & \widetilde{\eps}_5 :=
(e_{22}-e_{33})\otimes \pi_6,\\\nonumber &\eps_6 :=
(e_{23}-e_{32})\otimes \pi_5, & \widetilde{\eps}_6 :=
(e_{23}-e_{32})\otimes \pi_6.
\end{align}
{\bf Kaluza-Klein reduction of type IIA theory.} The low energy
limit of type IIA superstring theory yields ten-dimensional type IIA
supergravity. In order to cancel the $\hat{C}_{(7)}$-tadpole, it
turns out to be convenient for our purposes to allow for a nonzero
$\hat{F}_{(0)}$. This effectively leads to massive type IIA SUGRA
with mass $m_0=\hat{F}_{(0)}$. The corresponding action in the
string frame is given by\footnote{We use hats to indicate that a
field is ten-dimensional, following the conventions of
\cite{Grimm:2004ua}. Note also that in our convention for the RR
fields we have an additional factor of $\sqrt{2}$.}
\begin{align}\label{eq:action}
S^{(10)}_{IIA,m_0}= & S_{kin}+S_{CS}+S_{O6}\\\nonumber = &
\frac{1}{2\kappa_{10}^2} \int d^{10}x \sqrt{-\hat{g}}\left(
e^{-2\hat{\phi}}(\hat{R} + 4
\partial_{\mu}\hat{\phi} \partial^{\mu} \hat{\phi} - \frac{1}{2}|\hat{H}_3^{\mathrm{tot}}|^2)
-(|\hat{F}_2|^2 +|\hat{F}_4|^2+m_0^2)\right)\\\nonumber & -
\frac{1}{2\kappa_{10}^2}\int \left( \hat{B}_{(2)} \wedge d
\hat{C}_{(3)} \wedge d \hat{C}_{(3)} + 2  \hat{B}_{(2)} \wedge d
\hat{C}_{(3)} \wedge \hat{F}^{\mathrm{bg}}_{(4)}+ \hat{C}_{(3)}
\wedge \hat{H}^{\mathrm{bg}}_{(3)} \wedge d \hat{C}_{(3)}\right.
\\\nonumber & \quad \qquad \left. -\frac{m_0}{3} \hat{B}_{(2)}
\wedge \hat{B}_{(2)} \wedge \hat{B}_{(2)} \wedge d \hat{C}_{(3)} +
\frac{m_0^2}{20}
 \hat{B}_{(2)} \wedge \hat{B}_{(2)} \wedge \hat{B}_{(2)} \wedge \hat{B}_{(2)}
\wedge \hat{B}_{(2)}\right)\\\nonumber & + 2 \mu_6 \int_{O6} d^7\xi
e^{-\hat{\phi}} \sqrt{-\hat{g}} \: -2 \sqrt{2} \mu_6 \int_{O6}
\hat{C}_{(7)},
\end{align}
where $2\kappa_{10}^2 = (2 \pi)^7 {\alpha^{\prime}}^4$, $\mu_6 = (2
\pi)^{-6}{\alpha^{\prime}}^{-7/2}$ and the field strengths are given
by
\begin{subequations}
\begin{align}
\hat{H}^{\mathrm{tot}}_{(3)}&=d \hat{B}_{(2)}+ \hat{H}^{\mathrm{bg}}_{(3)},\\
\hat{F}_{(2)}&= d  \hat{C}_{(1)}+m_0 \hat{B}_{(2)},\\
\hat{F}_{(4)}&= d  \hat{C}_{(3)}+ \hat{F}^{\mathrm{bg}}_{(4)}-
 \hat{C}_{(1)} \wedge  \hat{H}^{\mathrm{tot}}_{(3)}-\frac{m_0}{2}
 \hat{B}_{(2)} \wedge \hat{B}_{(2)}.
\end{align}
\end{subequations}
In the framework of standard Kaluza-Klein reduction, we expand the
ten-dimensional gauge potentials in terms of harmonic forms on the
internal space $Y=T^6/{\mathbb Z}_4$, namely
\begin{align}
& \hat{C}_{(1)}= A^0(x),\; \hat{B}_{(2)}= B_{(2)}(x)+b^A(x)\omega_A,
\quad A = 1,\ldots, h^{(1,1)}, \\\nonumber &
\hat{C}_{(3)}=C_{(3)}(x)+ A^A(x)\wedge \omega_A +
\xi^{K}(x)\alpha_{K} - \tilde{\xi}_{K}(x)\beta^{K}, \quad
K=0,\ldots, h^{(2,1)}.
\end{align}
where $b^A,\xi^K,\tilde{\xi}_K$ are scalars in four dimensions,
$A^0,A^A$ are four-dimensional one-forms and $B_{(2)}$ and $C_{(3)}$
are four-dimensional two- and three-forms respectively. The harmonic
$(1,1)$-forms $\omega_A$ form a basis of $H^{(1,1)}(Y)$ with dual
$(2,2)$-forms $\tilde{\omega}_A$, which constitute a harmonic basis
of $H^{(2,2)}(Y)$. Moreover, $(\alpha_K, \beta^L)\in H^{(3)}(Y)$
form a real, sympletic basis of harmonic 3-forms on $Y$ with
dimension $h^{(3)}=2 h^{(2,1)}+2$. The intersection numbers are
\begin{equation}
\int_Y \alpha_K \wedge \beta^L = \delta_K^L, \quad \int_Y \omega_A
\wedge \tilde{\omega}^B = \delta_A^B.
\end{equation}
{\bf Details of the orientifold projection.}
After modding out by the orientifold projection ${\mathcal O}$, we will
be left with an ${\mathcal N}=1$ supergravity action.
To determine the ${\mathcal O}$-invariant states, first recall that the
ten-dimensional fields show the following behavior under $(-1)^{F_L}$ and
$\Omega_p$ (for a review, cf.~\cite{Grimm:2005fa}),
\begin{align}
(-1)^{F_L}&:\; \mathrm{odd:}\; \hat{C}_{(1)}, \hat{C}_{(3)}, \quad
\mathrm{even:}\; \hat{\phi},\hat{g},\hat{B}_{(2)},\\
\Omega_p&:\; \mathrm{odd:}\; \hat{B}_{(2)}, \hat{C}_{(3)}, \quad
\mathrm{even:}\; \hat{\phi},\hat{g},\hat{C}_{(1)}.
\end{align}
Accordingly, states that are ${\mathcal O}$-invariant have to satisfy
\begin{align}\label{eq:orientinv}
& \sigma^{\ast} \hat{\phi} = + \hat{\phi},\; \sigma^{\ast} \hat{g} =
+ \hat{g}, \; \sigma^{\ast} \hat{B}_{(2)}= -
\hat{B}_{(2)},\\\nonumber & \sigma^{\ast} \hat{C}_{(1)}= -
\hat{C}_{(1)},\; \sigma^{\ast} \hat{C}_{(3)}= + \hat{C}_{(3)}.
\end{align}
Therefore we want to investigate how the cohomology groups split
into even and odd subspaces under the antiholomorphic involution
$\sigma$,
\begin{equation}\label{eq:Hsplit}
H^p(Y)=H^p_+(Y) \oplus H^p_-(Y).
\end{equation}
The relevant cohomology groups together with their basis elements
are summarized in Table~3\footnote{Note that the volume form on
$T^6/{\mathbb Z}_4$ is odd under $\sigma$.}.
\begin{table}[h]
\begin{center}
\begin{tabular}{| c || c | c| c | c | c | c |} \hline
   \rule[-0.3cm]{0cm}{0.9cm} cohomology group &  $\ H^{(1,1)}_+\ $ &
   $\ H^{(1,1)}_-\ $ & $\ H^{(2,2)}_+\ $ & $\ H^{(2,2)}_-\ $ & $\ H^{(3)}_+\ $ & $\ H^{(3)}_-\ $
   \\ \hline
   \rule[-0.3cm]{0cm}{0.8cm} dimension &  $h^{(1,1)}_+$  & $h^{(1,1)}_- $
                                       &  $h^{(1,1)}_-$  & $h^{(1,1)}_+$
                                       &  $h^{(2,1)}+1$  &  $h^{(2,1)}+1$
   \\ \hline
   \rule[-0.3cm]{0cm}{0.8cm} basis     & $\omega_\alpha$ & $\omega_a$
                                       & $\tilde \omega^a$ & $\tilde \omega^\alpha$
   & $ a_{K}$ & $b^{K}$ \\ \hline
\end{tabular}
 \centerline{ \hbox{{\bf
Table 3:}{\it ~~Cohomology groups and their basis elements.}}}
\end{center}
\end{table}
\noindent
Let us begin by studying the behavior of the $(1,1)$-forms in the untwisted
sector. We will discuss the twisted sector moduli in chapter~\ref{sec:twisted}.
There are four $\sigma$-odd ${\mathbb Z}_4$-invariant (unnormalized)
$(1,1)$-forms, namely
\begin{subequations}
\begin{align}
&\sigma: (dz^i \wedge d\overline{z}^i) \mapsto -(dz^i \wedge
d\overline{z}^i), \quad i=1,2,3,\\
&\sigma: (dz^1 \wedge d\overline{z}^2+e^{\ic \pi/2}d\overline{z}^1
\wedge dz^2) \mapsto -(dz^1 \wedge d\overline{z}^2+e^{\ic
\pi/2}d\overline{z}^1 \wedge dz^2)
\end{align}
\end{subequations}
and one even $(1,1)$-form,
\begin{equation}
\sigma: (dz^1 \wedge d\overline{z}^2-e^{\ic \pi/2}d\overline{z}^1
\wedge dz^2) \mapsto +(dz^1 \wedge d\overline{z}^2-e^{\ic
\pi/2}d\overline{z}^1 \wedge dz^2).
\end{equation}
Consequently, $h_{+,untw.}^{(1,1)}=1$ and $h_{-,untw.}^{(1,1)}=4$.
Moreover, we can combine the ${\mathbb Z}_4$-invariant $(2,1)$-form and the
corresponding $(1,2)$-form into an even and an odd combination under $\sigma$,
\begin{equation}
\sigma: (dz^1\wedge dz^2 \wedge d\overline{z}^3 \pm \ic d\overline{z}^1 \wedge
d\overline{z}^2 \wedge dz^3)\mapsto \pm (dz^1\wedge dz^2 \wedge d\overline{z}^3
\pm \ic d\overline{z}^1 \wedge d\overline{z}^2 \wedge dz^3).
\end{equation}
{\bf Fluxes.} The following background fluxes of the NS-NS and R-R field
strengths are consistent with the orientifold projection and may thus be
turned on:
\begin{equation}\label{eq:fluxes}
\hat{F}^{\text{bg}}_0=m_0, \;
\hat{F}^{\text{bg}}_2=-m_{a}\omega_{a}, \;
\hat{F}^{\text{bg}}_4=e_{a}\tilde{\omega}^{a}, \;
\hat{H}^{\text{bg}}_3= - p_{K} b^{K},
\end{equation}
where we have taken into account the appropriate behavior of the
fluxes under $\sigma$. The indices $a= 1,\ldots,
h^{(1,1)}_{-,untw.}=4$ and $K=0,\ldots, h^{(2,1)}_{untw.}=1$ label
the basis elements of the cohomology groups, as given in table 3,
but are restricted to the untwisted sector. More explicitly, we have
\begin{subequations}
\begin{align}
&\omega_{1}= \left(\frac{\kappa}{2}\right)^{1/3}
\ic dz^{1}\wedge d\overline{z}^{1}, \\
&\omega_{2}= \left(\frac{\kappa}{2}\right)^{1/3}
\ic dz^{2}\wedge d\overline{z}^{2}, \\
&\omega_{3}= \left(\frac{\kappa}{2}\right)^{1/3}\frac{1}{U_2}
\ic dz^{3}\wedge d\overline{z}^{3}, \\
&\omega_4 =  \left(\frac{\kappa}{2}\right)^{1/3}\frac{(1-\ic)}{2}
(dz^1 \wedge d\overline{z}^2-\ic dz^2 \wedge d\overline{z}^1),
\end{align}
\end{subequations}
and in addition,
\begin{subequations}
\begin{align}
&\tilde{\omega}^{1}= \left(\frac{1}{(4\kappa)^{1/3}U_2}\right)
(\ic dz^{2}\wedge d\overline{z}^{2}) \wedge
(\ic dz^{3}\wedge d\overline{z}^{3}), \\
&\tilde{\omega}^{2}= \left(\frac{1}{(4\kappa)^{1/3}U_2}\right)
(\ic dz^{3}\wedge d\overline{z}^{3}) \wedge
(\ic dz^{1}\wedge d\overline{z}^{1}), \\
&\tilde{\omega}^{3}= \left(\frac{1}{(4\kappa)^{1/3}}\right)
(\ic dz^{1}\wedge d\overline{z}^{1}) \wedge
(\ic dz^{2}\wedge d\overline{z}^{2}), \\
&\tilde{\omega}^4 =- \left(\frac{1}{(4\kappa)^{1/3}U_2}\right)
\frac{(1-\ic)}{2} (dz^1 \wedge d\overline{z}^2-\ic dz^2 \wedge
d\overline{z}^1)\wedge (\ic dz^3 \wedge d\overline{z}^3),
\end{align}
\end{subequations}
such that
\begin{equation}
\int_Y \omega_1 \wedge \omega_2 \wedge \omega_3 = - \int_Y \omega_3
\wedge \omega_4 \wedge \omega_4 = \kappa
\end{equation}
and
\begin{equation}
 \int_Y \omega_{a}\wedge \tilde{\omega}^{b} = \delta_{a}^{b}.
\end{equation}
We normalize the volume form such that
\begin{equation}
\ic \int_Y \Omega \wedge \overline{\Omega} = 1 \; \Longrightarrow \; \Omega =
\frac{(1-\ic)}{2 \sqrt{U_2}} dz^1 \wedge dz^2 \wedge dz^3,
\end{equation}
and choose our three forms to be
\begin{subequations}
\begin{align}
&a_0=\frac{1}{2} (dx^1 \wedge dx^2 - dy^1 \wedge dy^2 + dx^1
\wedge dy^2 + dy^1 \wedge dx^2) \wedge dx^3, \\
&a_1 =\frac{1}{4} (dx^1 \wedge dx^2 - dy^1 \wedge dy^2 - dx^1
\wedge dy^2 - dy^1 \wedge dx^2) \wedge dy^3, \\
&b_0=2 (dx^1 \wedge dx^2 - dy^1 \wedge dy^2 + dx^1
\wedge dy^2 + dy^1 \wedge dx^2) \wedge dy^3, \\
&b_1 =-4 (dx^1 \wedge dx^2 - dy^1 \wedge dy^2 - dx^1 \wedge dy^2 -
dy^1 \wedge dx^2) \wedge dx^3.
\end{align}
\end{subequations}
$\Omega$ is given in this basis by
\begin{equation}
\label{eq:Omega} \Omega = \frac{1}{\sqrt{U_2}} \: a_0 +2
\sqrt{U_2} \: a_1 + \ic \frac{\sqrt{U_2}}{4} \: b_0 + \ic
\frac{1}{8 \sqrt{U_2}} \: b_1.
\end{equation}
The mixed-index part of the metric will be parameterized in the
following way,
\begin{equation}\label{eq:metric}
g_{i\bar{j}}=\left(\begin{matrix} \gamma_1 & \gamma_4 + \ic \gamma_5 & 0 \\
\gamma_4- \ic \gamma_5 & \gamma_2 &0 \\ 0 & 0  & \gamma_3 \end{matrix}\right).
\end{equation}
Taking into account the action of $\sigma$ on $g$, one finds that
$g_{1\bar{2}}=\ic g_{2\bar{1}}$, so that $\gamma_4=\gamma_5$.
Therefore, one K{\"a}hler modulus of the untwisted sector gets
projected out by the orientifold.\footnote{Note that there is a
non-vanishing metric component of pure type, namely
\begin{equation}
\delta g_{\bar{3}\bar{3}} = -\frac{1}{||\Omega||^2}
{\overline{\Omega}_{\bar{3}}}^{kl}(\chi_K)_{kl\bar{3}}(\tilde{z}^K),
\end{equation}
corresponding to the deformations of the complex structure. In our
conventions, the untwisted complex structure modulus $U_2$ also
shows up in the effective potential for the untwisted K{\"a}hler
moduli below.}

\section{Moduli stabilization}\label{sec:moduli}
We are now ready to calculate the potential for the various moduli
fields discussed above. In the next subsection, we will directly
calculate the potential from the (massive) IIA supergravity action
compactified on the orientifold in the presence of fluxes. Moreover,
we will derive several conditions, such as a tadpole cancelation
condition and another condition on the 3-form axions $\xi^0$ and
$\xi^1$ which are related to the complex structure.

\subsection{Dimensional (Kaluza-Klein) reduction from 10 to 4 dimensions}
Again, we shall first restrict ourselves to the untwisted sector of the
orientifold model.\\
\noindent {\bf Quantization of fluxes.} We impose the usual
cohomological quantization condition for a canonically normalized
field strength,
\begin{equation}\label{eq:quant}
\int \hat{F}_p = 2 \kappa_{10}^2 \mu_{8-p} f_p = (2 \pi)^{p-1}
\alpha'^{(p-1)/2}f_p.
\end{equation}
Accordingly, we have\footnote{Note the additional factor of
$\sqrt{2}$ for the RR fields in our conventions.}
\begin{equation}
m_0=\frac{f_0}{2\sqrt{2}\pi \sqrt{\alpha'}}, \;m_a = \frac{2\pi
\sqrt{\alpha'} f^{(a)}_2}{\sqrt{2}}, \; p_{K} = (2\pi)^2\alpha'
h_3^{(K)}, \; e_a = \frac{\kappa^{1/3}}{\sqrt{2}} (2\pi
\sqrt{\alpha'})^3 f^{(a)}_4,
\end{equation}
where $f_0,f^{(a)}_2,h_3^{(K)},f^{(a)}_4 \in {\mathbb Z}$.\\
\noindent {\bf Tadpole cancelation conditions.} The $O6$-plane will
generate a tadpole for the $\hat{C}_7$-potential, which we want to
cancel solely by background fluxes without adding $D6$-branes.
Noting that $\ast \hat{F}_{(2)}=d \hat{C}_7 -\hat{C}_5 \wedge
\hat{H}_3 - \frac{m_0}{24} \hat{B}_2 \wedge \hat{B}_2 \wedge
\hat{B}_2 \wedge \hat{B}_2$ contains $\hat{C}_7$, the integrated
equations of motion for the $\hat{C}_7$-potential yield
\begin{equation}
\label{eq:tad} \int d \hat{F}_{(2)} = \int m_0
\hat{H}^{\text{bg}}_3 \stackrel{!}{=} 2\sqrt{2}\kappa_{10}^2\mu_6
= 2 (\sqrt{2} \pi \sqrt{\alpha'}).
\end{equation}
The $O6$-plane wraps each of the cycles $[a_0]= (\rho_1 +\rho_2)$
and $[a_1]=(2 (\tilde{\rho}_1 - \tilde{\rho}_2) +\rho_2 - \rho_1)$
once. Thus we have to integrate (\ref{eq:tad}) over $[b_K], K=0,1$
leading to
\begin{equation}
m_0 p_K = -2 (\sqrt{2} \pi \sqrt{\alpha'}), \quad K=0,1.
\end{equation}
Taking into account the quantization condition (\ref{eq:quant}), we
arrive at the tadpole cancelation conditions
\begin{align}\label{eq:tadpole}
& m_0 p_0= m_0 p_1= (\sqrt{2} \pi \sqrt{\alpha'})f_0 h_3^{(K)} =-2
(\sqrt{2} \pi \sqrt{\alpha'}) \\\nonumber & \qquad \Rightarrow
(f_0,h_3^{(K)})=\pm (2,-1) \; {\text{or}} \; \pm (1,-2).
\end{align}
For later convenience we define $p \equiv p_0 =p_1$.\\
\noindent {\bf Potential for the untwisted complex structure axion.}
We will begin our discussion of the complex structure moduli by
considering the associated axions first. A more detailed examination
of the complex structure deformations will be carried out in the
next subsection. It actually turns out that the contribution to the
superpotential coming from $\hat{H}^{\text{bg}}_{(3)}$ fixes the
real part of the complex structure hypermultiplet (namely the
geometric complex structure moduli), while it leaves the imaginary
part (the axions) unfixed. After the orientifold projection, the
remaining axionic modes are\footnote{ We have chosen a symplectic
basis for $H^{(3)}(Y)$ such that all the $a_K$ are $\sigma$-even and
all the $b^K$ are $\sigma$-odd.}
\begin{equation}
\hat{C}_{(3)}= \xi^0 a_0 + \xi^1 a_1,
\end{equation}
noting that $\hat{C}_{(3)}$ has to be even under the involution
$\sigma$ in our construction. The discussion here mostly parallels
\cite{DeWolfe:2005uu}. The RR field $\hat{C}_{(3)}$ only appears in
the Chern-Simons piece of the massive IIA SUGRA action
({\ref{eq:action}). It is important to notice that $  \hat{C}_{(3)}
\wedge \hat{H}^{\mathrm{bg}}_{(3)}\wedge d \hat{C}_{(3)}$ is
nonvanishing only if $ d \hat{C}_{(3)}$ is polarized in the
noncompact directions. Since it does not contain physical degrees of
freedom, we will treat it as a Lagrange multiplier ${\mathcal F}_0
:= d C_{(3)}$. Plugging its equation of motion back into the action
yields
\begin{equation}
S_{{\mathcal F}_0 }= -\frac{1}{2 \kappa_{10}^2} \int {\mathcal F}_0 \wedge
\ast {\mathcal F}_0.
\end{equation}
Minimizing this contribution to the potential is tantamount to setting
${\mathcal F}_0=0$. Doing this and integrating over $Y$ results in an
equation involving the 3-form axions, namely
\begin{equation}\label{eq:axion}
p_0 \xi^0 + p_1 \xi^1 = e_0 +e_a b_a -\kappa m_0 b_3 (b_1 b_2 -
\frac{b_4^2}{2}),
\end{equation}
with the definition $e_0:=\int \hat{F}_{(6)}^{\text{bg}}$. This
means that only one linear combination of the axions is fixed
while there is another (independent) one that remains unfixed.
This is consistent with the results obtained below from analyzing
the superpotential. One could either try to stabilize the
remaining axion by introducing nonperturbative effects such as
Euclidean $D2$-instantons or by using the unfixed axion(s)
to give mass to (potentially anomalous) U(1) brane fields
via the St{\"u}ckelberg mechanism \cite{Camara:2005dc}. \\
\noindent {\bf Equations of motions for the $b_a$.} For simplicity
we will set\footnote{Solutions with $\hat{F}_2^{bg}\neq0$ have
qualitatively the same behavior as the $\hat{F}_2^{bg}=0$ solution
as will be shown later.} $\hat{F}_2^{bg}=0$. Since $\hat{C}_1$ has
no zero modes, the contributions from the $|\hat{F}_2|^2$ and
$|\hat{F}_4|^2$ terms in the action are at least quadratic in the
$b_a$. Since the Chern-Simons term linear in $\hat{B}_2$ has been
taken into account above, we find that the action contains no terms
linear in $b_a$. Therefore there is a solution with $b_a=0,\forall
a$. Since we will find supersymmetric and non-supersymmetric vacua
some of these solutions might have instabilities. We will further
investigate this at the end of this section. \\
\noindent {\bf Flux generated potential for the untwisted K{\"a}hler
and complex structure moduli.} In this section we will stabilize the
remaining untwisted moduli. We will work in the four dimensional
Einstein frame, so we define $g_{(4) \mu \nu} =
\frac{e^{\hat{\phi}}}{\sqrt{vol_{(6)}}} g_{(4)\mu \nu}^E$. The
effective potential is defined as
\begin{equation}
S= \frac{1}{\kappa_{10}^2}\int d^4x \sqrt{-g_{(4)}^E}
(-V_{\text{eff}}).
\end{equation}
For $b_a=0$ and the $\xi^K$ satisfying their equation of motion we
only get contributions from the terms $|\hat{H}_3^{\mathrm{tot}}|^2,
|\hat{F}_4|^2, m_0^2$ and the $O6$ Born-Infeld piece. They are
\begin{align}
\nonumber V_{\text{eff}}= & \frac{e^{2\hat{\phi}}}{vol_{(6)}^2} p^2
(\frac{1}{U_2}+4 U_2)+ \frac{e^{4\hat{\phi}}}{2 vol_{(6)}^3}\left[
\sum_{i=1}^{3}e_i^2v_i^2 + e_4^2(v_1 v_2 +\frac{v_4^2}{2}) + e_1 e_2
v_4^2+ 2 e_4 v_4 (e_1 v_1 + e_2 v_2) \right]\\ & +\frac{m_0^2}{2}
\frac{e^{4\hat{\phi}}}{vol_{(6)}}- 2 |m_0 p|
\frac{e^{3\hat{\phi}}}{vol_{(6)}^{3/2}} \left(
\frac{1}{\sqrt{U_2}}+2 \sqrt{U_2} \right),
\end{align}
where
\begin{align*}
v_1 & = \frac{1}{2}\left( \frac{2}{\kappa}\right)^{1/3} \gamma_1,
\qquad \qquad
v_2  = \frac{1}{2}\left( \frac{2}{\kappa}\right)^{1/3} \gamma_2,\\
v_3 & = \frac{1}{2}\left( \frac{2}{\kappa}\right)^{1/3} U_2 \:
\gamma_3, \quad \quad \: \: \;
v_4  = - \left( \frac{2}{\kappa}\right)^{1/3} \gamma_4,\\
vol_{(6)} & = \int_Y dx^1 \wedge dy^1 \wedge dx^2 \wedge dy^2
\wedge dx^3 \wedge dy^3 \sqrt{g_{(6)}} = U_2 \frac{\gamma_3
(\gamma_1 \gamma_2 -2 \gamma_4^2)}{4}= \kappa v_3 (v_1 v_2 -
\frac{v_4^2}{2}).
\end{align*}
Extremizing the potential with respect to the complex structure
$U_2$ fixes it at
\begin{equation}
U_2 = \frac{1}{2}.
\end{equation}
Now we solve
\begin{equation}
v_a \frac{\partial V}{\partial v_a}+ \frac{7}{4} \frac{\partial
V}{\partial \hat{\phi}}=0,
\end{equation}
and find
\begin{equation}\label{eq:const}
e^{\hat{\phi}}\sqrt{vol_{(6)}}=\frac{5}{\sqrt{2}} \left|
\frac{p}{m_0} \right|.
\end{equation}
This is (almost) fixed by the tadpole cancelation conditions,
cf.~equation~(\ref{eq:tadpole}) above. This condition ensures that,
for minima of the potential, the string coupling automatically
becomes small if we tune the fluxes such that the internal volume
becomes large enough to trust the supergravity approximation we are
using. Relation (\ref{eq:const}) can be used to eliminate the
dilaton dependence of the potential. Once the minima have been
found, said relation fixes the dilaton w.r.t. a specific set of
fluxes. The potential simplifies to
\begin{equation}
\nonumber V_{\text{eff}}= \frac{25}{8} \frac{p^4}{m_0^2}
\left(\frac{-39}{vol_{(6)}^3}+ \frac{25}{m_0^2 \, vol_{(6)}^5}
\left[ \sum_{i=1}^{3}e_i^2v_i^2 + e_4^2(v_1 v_2 +\frac{v_4^2}{2}) +
e_1 e_2 v_4^2+ 2 e_4 v_4 (e_1 v_1 + e_2 v_2) \right] \right).
\end{equation}
It now only depends on the K\"ahler moduli $v_1, v_2,v_3,v_4$.
Extremizing with respect to all of the K\"ahler moduli leads to five
sets of solutions. The first is
\begin{align}
\label{eq:solution1} v_1 &= \pm e_2 \sqrt{\frac{10}{3}} \sqrt{\left|
\frac{e_3}{ \kappa m_0(2 e_1 e_2 - e_4^2)}\right|},\\\nonumber v_2
&= \pm e_1 \sqrt{\frac{10}{3}} \sqrt{\left| \frac{e_3}{ \kappa m_0(2
e_1 e_2 - e_4^2)}\right|},\\\nonumber v_3 &= \sqrt{\frac{5}{6}}
\sqrt{\left| \frac{2 e_1 e_2 - e_4^2}{\kappa m_0 e_3}
\right|},\\\nonumber v_4 &= \mp e_4 \sqrt{\frac{10}{3}} \sqrt{\left|
\frac{e_3}{ \kappa m_0(2 e_1 e_2 - e_4^2)}\right|}.
\end{align}
As we will see below this solution encompasses the supersymmetric
solution obtained from minimizing the potential of the 4-d SUGRA
action. To allow for a geometrical interpretation of the solution we
have to demand that the volume $vol_{(6)}$ and $v_3$ the area of the
third torus are bigger than zero. This implies that $(2 e_1 e_2
-e_4^2)>0$ which requires $\sign[e_1 e_2]>0$. The volume is
\begin{equation}
vol_{(6)}=\frac{5}{3} \sqrt{\frac{5}{6}} \sqrt{\left| \frac{e_3 (2
e_1 e_2 -e_4^2)}{\kappa m_0^3} \right|}.
\end{equation}
It can be made parametrically large by tuning the fluxes to large
values. The string coupling is determined to be
\begin{equation}
g_s= e^{\hat{\phi}}= |p| \left( \frac{135}{2} \left|
\frac{\kappa}{m_0 e_3 (2 e_1 e_2-e_4^2)} \right| \right)^{1/4}.
\end{equation}
Thus, there is a (countably) infinite number of vacua with small
string coupling and large volume.\footnote{If we set for example
$e_1=e_2=e_3=e_4\equiv e\rightarrow \infty$, we have $vol \sim
e^{3/2}, \quad e^{\hat{\phi}} \sim e^{-3/4}$.} \\
The value of the potential at the minimum is
\begin{equation}
V_{\text{min}}=-\frac{243}{25} \sqrt{\frac{6}{5}} \sqrt{\left|
\frac{\kappa^3 m_0^5}{(e_3 (2 e_1 e_2 -e_4^2))^3} \right|} \: p^4,
\end{equation}
which is always negative so that the vacua are anti-de-Sitter.\\
The second set of solutions is
\begin{align}
v_1 &= \pm e_4 \sqrt{\frac{5}{3}} \sqrt{\left| \frac{e_2 e_3}{
\kappa m_0 e_1 (2 e_1 e_2 - e_4^2)}\right|},\\\nonumber v_2 &= \pm
e_4 \sqrt{\frac{5}{3}} \sqrt{\left| \frac{e_1 e_3}{ \kappa m_0 e_2
(2 e_1 e_2 - e_4^2)}\right|},\\\nonumber v_3 &= \sqrt{\frac{5}{6}}
\sqrt{\left| \frac{2 e_1 e_2 - e_4^2}{\kappa m_0 e_3}
\right|},\\\nonumber v_4 &= \mp 2 \sqrt{\frac{5}{3}} \sqrt{\left|
\frac{e_1 e_2 e_3}{ \kappa m_0(2 e_1 e_2 - e_4^2)}\right|}.
\end{align}
For this case we have to demand that $(2 e_1 e_2 -e_4^2)<0$ and
$\sign[e_1 e_2]>0$. The volume, the string coupling and the
potential at the minimum are the same as above. This is also the
case for all the other solutions.\\
The next set of solutions has $v_4$ fixed at zero
\begin{align}
v_1 &= \pm \sqrt{\frac{5}{3}} \sqrt{\left| \frac{e_2 e_3}{ \kappa
m_0 e_1}\right|},\\\nonumber v_2 &= \pm \sqrt{\frac{5}{3}}
\sqrt{\left| \frac{e_1 e_3}{ \kappa m_0 e_2}\right|},\\\nonumber v_3
&= \sqrt{\frac{5}{6}} \sqrt{\left| \frac{2 e_1 e_2 - e_4^2}{\kappa
m_0 e_3} \right|},\\\nonumber v_4 &= 0.
\end{align}
It requires $\sign[e_1 e_2]<0$ which implies $(2 e_1 e_2
-e_4^2)<0$.\\
We furthermore find solutions in which one of the K\"ahler moduli is
unstabilized
\begin{align}
v_1 &= \frac{1}{e_1^2} \left( (-e_1 e_2 +e_4^2) v_2 \pm e_4 \sqrt{
|2 e_1 e_2 -e_4^2| v_2^2 - \left| \frac{10}{3} \frac{e_1^2
e_3}{\kappa m_0} \right| } \right),\\\nonumber v_2 &= {\text {\bf
unfixed}},\\\nonumber v_3 &= \sqrt{\frac{5}{6}} \sqrt{\left| \frac{2
e_1 e_2 - e_4^2}{\kappa m_0 e_3} \right|},\\\nonumber v_4 &=
\frac{1}{e_1} \left( (-e_4 v_2 \mp \sqrt{ |2 e_1 e_2 -e_4^2| v_2^2 -
\left| \frac{10}{3} \frac{e_1^2 e_3}{\kappa m_0} \right| } \right).
\end{align}
These solutions require $(2 e_1 e_2 -e_4^2)<0$ and $v_2^2> \left|
\frac{10}{3} \frac{e_1^2 e_3}{\kappa m_0 (2 e_1 e_2 -e_4^2)}
\right|$. Since the action is invariant under the simultaneous
exchange of $e_1 \leftrightarrow e_2$ and $v_1 \leftrightarrow v_2$,
we have corresponding solutions in which $v_1$ is unfixed.\\
Although we have turned on the most generic fluxes compatible with
the orbifold and orientifold projection, we found solutions that
have one unstabilized geometric modulus. As we will see below these
solutions are not supersymmetric.\\
\noindent {\bf Stability analysis for the $b_a$.} Since we have
found vacua that are non-supersymmetric, we have to check that our
$b_a=0$ solution is in fact stable. To do this we consider the terms
quadratic in $b_a$ and $\xi^K$ \footnote{Remember that
(\ref{eq:axion}) implies that there is a mixing between the $b_a$
and $\xi^K$.}. We find
\begin{align} S_{\text axion} &= \frac{1} {2 \kappa_{10}^2} \int
d^4x \sqrt{-g_4^E} \left[ -\frac{1}{2 vol_{(6)}} \partial_\mu b^a
\partial^\mu b^b \int_Y (\omega_{a} \wedge \ast_6 \, \omega_{b}) - e^{2D} \partial_\mu
\xi^K \partial^\mu \xi^L \int_Y (a_{K} \wedge \ast_6 \, a_{L})
\right.
\\\nonumber &- \left. e^{4D} \left( m_0^2 b^a b^b \int_Y (\omega_{a} \wedge \ast_6
\, \omega_{b}) - m_0 b^a b^b e_c \int_Y (\omega_{a} \wedge
\omega_{b} \wedge \ast_6 \,\tilde{\omega}^{c}) +\frac{(-p_K \xi^K +
e_a b^a)^2} {vol_{(6)}} \right) \right],
\end{align}
where we defined the four dimensional dilation as
$e^D=\frac{e^{\hat{\phi}}}{\sqrt{vol_{(6)}}}$. Now one has to
diagonalize the kinetic energy terms and calculate the mass-squared
matrix (Hessian) for each of the solutions described above.
To carry out the calculations in full
generality is rather tedious. From the action we see that the result
will depend on the explicit choices for the fluxes $m_0$ and $e_a$.
We have calculated the mass-squared matrix for simple sets of fluxes
for all of our vacua. In each case, we obtain positive mass eigenvalues
with the exception of one zero eigenvalue corresponding to the unstabilized
axion $\xi_0 - \xi_1$ (cf. (\ref{eq:axion})). Thus, there exists
a stable solution for all vacua (with large fluxes).
In conclusion, we see that the solution corresponding to
$b_a=0,\; \forall a$,
is a stable minimum of the effective four-dimensional potential, at
least for simple choices of the fluxes.\\

\subsection{Effective ${\mathcal N}=1$ SUGRA in $D=4$}
In this subsection we will analyze the problem from the point of
view of the effective ${\mathcal N}=1$ SUGRA theory in four
dimensions. One of the virtues of working in this framework is that
the untwisted and the twisted moduli can be treated on equal
footing. As pointed out in \cite{DeWolfe:2005uu}, another advantage
lies in the fact that this type of analysis can be used for general
backgrounds since e.g., backreaction and worldsheet instanton
corrections are naturally described in terms of the four-dimensional
effective theory, whereas they cannot be described in terms of
ten-dimensional supergravity. Based on the flux-generated
superpotential, as worked out by Grimm and Louis \cite{Grimm:2004ua}
(see also \cite{Grimm:2005fa}), we will analyze the F-flatness
conditions $D_I W=0$, where $I$ runs over all moduli fields and $D_I
= \partial_I + (\partial_I K)$ is the K{\"a}hler covariant
derivative. Solutions to these equations correspond to
supersymmetric minima of the scalar potential,
\begin{equation}
V=e^{K}\left(\sum_{I\bar{J}}G^{I\bar{J}}D_I W \overline{D_J W}-3 |W|^2 \right)
+m_0 e^{K^Q}\text{Im}W^Q,
\end{equation}
namely
\begin{equation}\label{eq:Fflat}
D_I W = 0 \Rightarrow dV=0.
\end{equation}
The opposite direction is not true. The structure of the K{\"a}hler
potential $K=K^K+K^Q$ and the superpotential $W=W^K+W^Q$ will be
discussed below.\\
\noindent {\bf ${\mathcal N}=2$ SUGRA in $D=4$.} The dimensional
reduction of (massive) type IIA supergravity from $D=10$ to $D=4$ on
a Calabi-Yau manifold gives rise to ${\mathcal N}=2$ supergravity in
$D=4$. The existence of one covariantly constant spinor on the
internal CY (with $SU(3)$ holonomy) ensures that there are two
four-dimensional SUSY parameters; the compactification therefore
preserves eight supercharges, hence ${\mathcal N}=2$ in $D=4$. In
the presence of fluxes, the resulting effective theory in four
dimensions is gauged, i.e., the hypermultiplets are charged under
some of the vectormultiplets. For this to be consistent, the metric
on the scalar manifold coordinatized by the hypermultiplets, which
is in fact a quaternionic manifold, must possess isometries that in
turn can be gauged. Table 4 lists the bosonic components of all
${\mathcal N}=2$ multiplets.
\begin{table}[h]
\begin{center}
\begin{tabular}{| c | c | c |} \hline
   \rule[-0.3cm]{0cm}{0.8cm} gravity multiplet  &
   $1$ & {\small $(g_{\mu \nu},A^0)$}
   \\ \hline
   \rule[-0.3cm]{0cm}{0.8cm} vectormultiplets &
   $h^{(1,1)}$ & {\small $(A^{A}, v^A,b^A)$}\\ \hline
   \rule[-0.3cm]{0cm}{0.8cm} hypermultiplets  &
   $h^{(2,1)}$ &
   {\small $(z^K,\xi^K,\tilde \xi_K)$}\\ \hline
\rule[-0.3cm]{0cm}{0.8cm} tensor multiplet  &
   1 &
   {\small $(B_{(2)},\hat{\phi},\xi^0,\tilde \xi_0)$} \\ \hline
\end{tabular}
\centerline{\hbox{{\bf
Table 4:}{\it ~~ Bosonic part of the ${\mathcal N}=2$ multiplets for Type IIA SUGRA on a CY3.}}}
\end{center}
\end{table}
\noindent
There are massless modes coming from deformations of the metric $g$
of the CY manifold that respect the Ricci flatness condition $R_{mn}=0$. This
forces $\delta g$ to satisfy the Lichnerowicz equation, whose solutions in our
case can be identified with harmonic (1,1)- and (2,1)-forms on $Y$,
corresponding to K{\"a}hler structure and complex structure deformations,
respectively. \\
\noindent {\bf K{\"a}hler moduli space.} Deformations of the
K{\"a}hler form can be expanded in a basis of harmonic (1,1)-forms,
\begin{equation}
g_{i\bar{j}}+\delta g_{i\bar{j}}= - \ic J_{i\bar{j}}=-\ic
v^A(\omega_A)_{i\bar{j}}, \quad A=1,\ldots , h^{(1,1)}.
\end{equation}
These deformations can be supplemented by the $h^{(1,1)}$ real scalar fields
$b^A(x)$ from the expansion of the B-field, yielding complex fields
\begin{equation}
t^A=b^A+\ic v^A,
\end{equation}
that parametrize the complexified K{\"a}hler cone. The moduli space
of the complexified K{\"a}hler structure deformations ${\mathcal
M}^{\text{ks}}$ is a special K{\"a}hler manifold which can be seen
by noting that the metric is given by
\begin{equation}
G_{AB}=\frac{3}{2 \kappa}\int_Y \omega_A \wedge \ast \omega_B =
-\frac{3}{2} \left(\frac{\kappa_{AB}}{\kappa}-\frac{3}{2}
\frac{\kappa_A \kappa_B}{\kappa^2}\right)=
\partial_{t^A}\partial_{t^B}K^{\text{ks}},
\end{equation}
where the intersection numbers are defined as follows
\begin{align*}
\kappa &= \int_Y J \wedge J \wedge J = \kappa_{ABC} v^A v^B v^C,\quad
\kappa_A = \int_Y \omega_A \wedge J \wedge J = \kappa_{ABC} v^B v^C,\\
\kappa_{AB} &= \int_Y \omega_A \wedge \omega_B \wedge J = \kappa_{ABC} v^C,
\quad \kappa_{ABC} = \int_Y \omega_A \wedge \omega_B \wedge \omega_C.
\end{align*}
The K{\"a}hler potential for the K{\"a}hler structure deformations,
\begin{equation}
K^{\text{ks}}= -\ln \left(\frac{\ic}{6}\kappa_{ABC}(t-\bar{t})^A(t-\bar{t})^B
(t-\bar{t})^C \right)=-\ln \frac{4}{3}\kappa,
\end{equation}
can be derived from a single holomorphic prepotential
${\mathcal G}(t)= -\frac{1}{6} \kappa_{ABC}t^At^Bt^C$.\\
\noindent {\bf Complex structure moduli space.} Complex structure
deformations are associated with harmonic (1,2)-forms and are
parametrized by complex fields $\tilde{z}^K$,
$K=1,\ldots,h^{(2,1)}$, in the following way,
\begin{equation}
\delta g_{ij} =\frac{\ic}{||\Omega||^2} \overline{\tilde{z}}^K
(\overline{\chi}_K)_{i\bar{i}\bar{j}} {\Omega^{\bar{i}\bar{j}}}_j,
\end{equation}
where the $\chi_K$ form a harmonic basis of $H^{(2,1)}(Y)$ and
$||\Omega||^2= \frac{1}{3!}\Omega_{ijk}\Omega^{ijk}$. The metric on
the complex structure moduli space ${\mathcal M}^{\text{cs}}$ is
given by
\begin{equation}
G_{K\bar{L}}=-\frac{\int_Y \chi_K \wedge \overline{\chi}_L}{\int_Y \Omega \wedge
\overline{\Omega}}.
\end{equation}
Kodaira's formula connects the $\chi_K$ to the variation of the
harmonic (3,0)-form via
\begin{equation}
\chi_K(\tilde{z}, \overline{\tilde{z}})= \partial_{\tilde{z}^K} \Omega
(\tilde{z}) +\Omega (\tilde{z}) \partial_{\tilde{z}^K}K^{\text{cs}},
\end{equation}
where
\begin{equation}
K^{\text{cs}}(\tilde{z}, \overline{\tilde{z}})= - \ln \left[ \ic
\int_Y \Omega \wedge \overline{\Omega} \right] =  - \ln \ic \left[
\overline{Z}^{K} {\mathcal F}_{K} - Z^{K} \overline{\mathcal
F}_{K}\right].
\end{equation}
Note that $G_{K\bar{L}}=
\partial_{\tilde{z}^K}\partial_{\tilde{z}^{\bar{L}}} K^{\text{cs}}$,
thus proving that that ${\mathcal M}^{\text{cs}}$ is a K{\"a}hler
manifold. The holomorphic periods $Z^{K}, {\mathcal F}_{K}$ are the
expansion coefficients of
\begin{equation}
\Omega = Z^{K} \alpha_{K} - {\mathcal F}_{K}\beta^{K},
\end{equation}
so that we have
\begin{equation}
Z^{K} = \int_Y \Omega \wedge \beta^{K}, \quad {\mathcal F}_{K}=
\int_Y \Omega \wedge \alpha_{K}.
\end{equation}
In fact, $\Omega$ is only defined up to a complex rescaling with a holomorphic
function which changes the K{\"a}hler potential by a K{\"a}hler transformation.
This symmetry can be used to fix a K{\"a}hler gauge, in which $Z^0=1$.
The remaining periods can be identified with the $h^{(2,1)}$ complex structure
deformations
\begin{equation}
\tilde{z}^K= \frac{Z^K}{Z^0}.
\end{equation}
Moreover, we find that there exists a prepotential of which
${\mathcal F}_{K}$ is the first derivative, ${\mathcal F}=
\frac{1}{2} Z^{K}{\mathcal F}_{K}$. This means that the metric
$G_{K\bar{L}}$ is completely determined by ${\mathcal F}$. Therefore
${\mathcal M}^{\text{cs}}$ is in fact a special K{\"a}hler manifold.\\
Supplementing the complex structure deformations $\tilde{z}^K$ with
the corresponding axions $\xi^K$ and $\tilde{\xi}_K$ from the RR
3-form $\hat{C}_3$ can be shown to result in a special quaternionic
structure of the resulting moduli space. We will refer to this
larger manifold, spanned by the scalars in the hypermultiplets, as
${\mathcal M}^Q$. In the next section we will use the fact that
${\mathcal M}^Q$ contains the special K{\"a}hler submanifold
${\mathcal M}^{\text{cs}}$ spanned by the complex
structure deformations.\\
\noindent {\bf Orientifold projection.} As already mentioned above,
the cohomology groups split into even and odd parts under the
antiholomorphic involution $\sigma$ (cf.~(\ref{eq:Hsplit})). The
involution must act as \cite{Grana:2005jc}
\begin{equation}
\sigma^{\ast} J= -J, \quad \sigma^{\ast} \Omega = e^{2\ic \theta}
\overline{\Omega}.
\end{equation}
The fixed loci of $\sigma$ (which the $O6$-plane wraps) are special
Lagrangian (sLag) 3-cycles $\Sigma_n$ fulfilling
\begin{equation}
J\Big|_{\Sigma_n} = 0, \quad {\text{Im}}(e^{-\ic \theta}\Omega)\Big|_{\Sigma_n}
= 0.
\end{equation}
Together with the conditions (\ref{eq:orientinv}) we are left with
\begin{equation}
J_c := B + \ic J = \sum_{a=1}^{h^{(1,1)}_-} t^a \omega_a.
\end{equation}
Thus, the orientifold projection reduces the K{\"a}hler moduli space
to a subspace without altering its complex structure and the
K{\"a}hler potential is inherited directly from ${\mathcal N}=2$,
\begin{equation}
K^K(t^a)= - \log (\frac{4}{3}\kappa_{abc} v^a v^b v^c).
\end{equation}
For the holomorphic (3,0)-form, we get
\begin{equation}
\Omega (\tilde{z}) = Z^{K}(\tilde{z}) a_{K} - {\mathcal
F}_{K}(\tilde{z})b^{K},
\end{equation}
where we have decomposed $H^{(3)}(Y)=H_+^{(3)}(Y)\oplus
H_-^{(3)}(Y)$ as indicated in table 3. As remarked upon earlier, one
can always perform a symplectic rotation on the resulting even and
odd bases such that all $a_{K}$ are even and all $b^{K}$ are odd.
Note that the $h^{(1,1)}_+$ vector multiplets do not contain any
scalars and will therefore be disregarded. It is customary to
package the remaining degrees of freedom in the following way,
\begin{equation}
\Omega_c = \hat{C}_{(3)} +2 \ic {\text{Re}}(C\Omega),
\end{equation}
where we have introduced the complex compensator $C=re^{-\ic
\theta}$, where $r=e^{-D+K^{\text{cs}/2}}$. $r$ transforms
oppositely to the holomorphic 3-form under holomorphic
transformations so as to render $C\Omega$ scale-invariant (the
compensator replaces the irrelevant scale factor in favor of the
physical dilaton field $D$; for more details see \cite{Grimm:2004ua,
Grana:2005jc}). The field $\hat{C}_{(3)}=\xi^{K}a_{K}$ comprises the
surviving axionic modes. Finally, $\Omega_c$ can be expanded in a
basis of $H^{(3)}_+(Y)$,
\begin{equation}
\Omega_c=2 N^{K} a_{K},
\end{equation}
where
\begin{equation}
N^{K}= \frac{1}{2}\int_Y \Omega_c \wedge b^{K}= \frac{1}{2}(\xi^{K}
+ 2 \ic {\text{Re}}(C Z^{K})).
\end{equation}
We have now reduced the number of moduli, while preserving the
original ${\mathcal N}=2$ complex structure. Table 5 shows the
surviving ${\mathcal N}=1$ spectrum.
\begin{table}[ht]
\begin{center}
\begin{tabular}{|l|c|c|} \hline
 \rule[-0.3cm]{0cm}{0.8cm}
multiplets& multiplicity & bosonic components\\ \hline\hline
 \rule[-0.3cm]{0cm}{0.8cm}
 gravity multiplet&1&$g_{\mu \nu} $ \\ \hline
 \rule[-0.3cm]{0cm}{0.8cm}
 vector multiplets&   $h_+^{(1,1)}$&  $A^{\alpha} $\\ \hline
 \rule[-0.3cm]{0cm}{0.8cm}
 {chiral multiplets} &   $h_-^{(1,1)}$&
$t^a$ \\ \hline
 \rule[-0.3cm]{0cm}{0.8cm}
{chiral multiplets}
& $ h^{(2,1)}+1$ &$ N^{K}$\\
\hline
\end{tabular}
\centerline{\hbox{{\bf Table 5:}{\it ~~ ${\mathcal N}=1$ multiplets
after orientifold projection.}}}
\end{center}
\end{table}
The ${\mathcal N}=1$ K{\"a}hler potential is given by
\begin{equation}
K^Q= -2 \log \left(2 \int {\text{Re}}(C\Omega) \wedge \ast {\text{Re}}(C\Omega)
\right) = 4D,
\end{equation}
where
\begin{equation}
 \int {\text{Re}}(C\Omega) \wedge \ast {\text{Re}}(C\Omega) =
- {\text{Re}}(C Z^{K}){\text{Im}}(C{\mathcal F}_{K}) =
\frac{e^{-2D}}{2}.
\end{equation}
For the four dimensional dilaton we have
\begin{equation}
e^D = \frac{e^{\hat{\phi}}}{\sqrt{vol}} = \sqrt{8} e^{\hat{\phi}
+K^K/2}.
\end{equation}
In conclusion, we have seen that from each quaternionic
hypermultiplet only the real part of the complex structure modulus
and one axion survives. The degrees of freedom in the universal
hypermultiplet are also cut in half, namely the dilaton $\hat{\phi}$
and the axion $\xi^0$ survive.

\subsection{Supersymmetric AdS vacua}

It was demonstrated by Grimm and Louis \cite{Grimm:2004ua} that
dimensionally reducing massive type IIA supergravity from 10 to 4
dimensions, while neglecting the backreaction of the fluxes and
other local sources on the geometry of the compactification
manifold, leads to the following scalar potential,
\begin{equation}
V= e^{K^K+K^Q} \left(\sum_{I,J=t^a,N^K}G^{I\bar{J}}D_I W
\overline{D_J W} - 3 |W|^2\right)+m_0 e^{K^Q}{\text{Im}}W^Q.
\end{equation}
The second term cancels with contributions from the $O6$-plane when
the tadpole cancelation condition (\ref{eq:tad}) is satisfied. The
superpotential is given by
\begin{subequations}
\begin{align}
W(t^a,N^{K})&= W^Q(N^{K})+W^K(t^a),\\
W^Q(N^{K})&= \int_Y \Omega_c \wedge \hat{H}_{(3)} = -2 p_{K} N^{K} =
-p_{K} \xi^{K} - 2 \ic
p_{K} {\text{Re}}(C Z^{K}),\\
W^K(t^a)&= \int_Y e^{-J_c} \wedge \hat{F} = e_0 +\int_Y J_c \wedge
\hat{F}_{(4)} -\frac{1}{2} \int_Y J_c \wedge J_c \wedge
\hat{F}_{(2)} -\frac{m_0}{6} \int_Y J_c \wedge J_c \wedge J_c
\\\nonumber &= e_0 +e_a t^a + \frac{1}{2}\kappa_{abc} t^a t^b m^c -
\frac{m_0}{6} \kappa_{abc} t^a t^b t^c,
\end{align}
\end{subequations}
with the definition $\hat{F}= m_0
-\hat{F}^{\text{bg}}_{(2)}-\hat{F}^{\text{bg}}_{(4)}
+\hat{F}^{\text{bg}}_{(6)}$ (cf.~(\ref{eq:fluxes})). In the
following sections we will first analyze the equations for the
moduli from the F-term conditions (\ref{eq:Fflat}) in general and
then specialize to the case at hand, namely the $T^6/{\mathbb Z}_4$
orientifold. It is important to note that these equations will be
valid for all (untwisted and twisted) moduli.
The discussion closely follows the one in \cite{DeWolfe:2005uu}.\\
\noindent {\bf Complex structure equations.} Solving for $D_{N^{K}}
W=0$ yields
\begin{equation}\label{eq:cmplx}
p_{K} + 2 \ic W {\text{Im}}(C {\mathcal F}_{K}) e^{2D}= 0.
\end{equation}
We shall study the real and imaginary parts of this equation separately.
For the real part one gets
\begin{equation}
p_{K} - 2 e^{2D} {\text{Im}}(W){\text{Im}}(C {\mathcal F}_{K})=0.
\end{equation}
We immediately learn from this equation that ${\text{Im}}(W)=0$ is
incompatible with non-vanishing $\hat{H}^{\text{bg}}_{(3)}$-flux.
Thus assuming $\text{Im}(W)\neq 0$ we find that for each
$p_{K_i}=0$, we have $\text{Im} (C {\mathcal F}_{K_i})=0$. For
$p_{K_j}\neq 0$, one finds
\begin{equation}
\label{eq:cstructure} e^{-K^{\text{cs}}/2}
\frac{p_{K_j}}{\text{Im}({\mathcal F}_{K_j})}= 2 e^D
\text{Im}(W)=:Q_0,
\end{equation}
thus fixing all geometric complex structure moduli (including the
twisted ones, in our case $K= 0, \ldots, h^{(2,1)}=7$). As noted
above, these equations are invariant under rescalings of $\Omega$
and therefore do only depend on the $h^{(2,1)}$ inhomogeneous
coordinates of ${\mathcal M}^{\text{cs}}$, yielding $h^{(2,1)}$
equations for the $h^{(2,1)}$ moduli. The dilaton will be stabilized
at
\begin{equation}
\label{eq:dilaton} e^{-\hat{\phi}}= 4 \sqrt{2} e^{K^K/2}
\frac{\text{Im}(W)}{Q_0},
\end{equation}
once complex structure and K{\"a}hler moduli are fixed.\\
Turning to the imaginary part of (\ref{eq:cmplx}), we see that, due
to the reality of the flux coefficients $p_{K}$, all ${K}$ equations
yield the same condition, namely ($D$ and $C=r$ are real\footnote{We
absorb $\theta$ in the holomorphic 3-form so that it satisfies
$\sigma^{\ast} \Omega = \overline{\Omega}$}.)
\begin{equation}
\label{eq:reW} 2 e^{2D}{\text{Re}}(W) {\text{Im}}(C{\mathcal F}_{K})
=0 \Rightarrow {\text{Re}}(W) =0.
\end{equation}
Comparing to the definition of $W$, this indeed gives the same condition on
the axions as derived above (cf.~(\ref{eq:axion})),
\begin{equation}
\label{eq:axion4d} -p_{K} \xi^{K} + {\text{Re}}(W^K) = 0,
\end{equation}
where we have now correctly considered all the axions, including
those from the twisted sectors. Another important observation can be
made by multiplying (\ref{eq:cmplx}) by ${\text{Re}}(C Z_K)$ and
summing over $K$. The resulting equation reads
\begin{equation}
-\ic W = - p_{K}{\text{Re}}(C Z^{K})= \frac{1}{2}\text{Im}(W^Q).
\end{equation}
Now since ${\text Re}(W)=0$ (cf.~(\ref{eq:reW})), we find
\begin{equation}
-\ic W = \text{Im}(W^K)+\text{Im}(W^Q)= \frac{1}{2}\text{Im}(W^Q)\\
\Rightarrow \text{Im}(W^Q)=- 2 \, \text{Im}(W^K).
\end{equation}
Therefore we can directly conclude that, provided the complex
structure moduli are `on-shell' (satisfy their equations of motion),
the vacuum superpotential can be given solely in terms of the
K{\"a}hler moduli, i.e.,
\begin{equation}\label{eq:newW}
W(t^a, N^{K})= -\ic \, \text{Im}(W^K(t^a)),
\end{equation}
thus effectively decoupling the K{\"a}hler sector from the complex structure
sector.\\
\noindent {\bf K{\"a}hler structure equations.} Let us now consider
the K{\"a}hler sector in more detail. The corresponding F-flatness
conditions $D_{t^a}W=0$ can be simplified making use of
(\ref{eq:newW}), yielding
\begin{equation}\label{eq:Kahlereq}
\partial_{t^a} W^K - \ic \partial_{t^a}K^K {\text{Im}}(W^K)=0.
\end{equation}
The imaginary parts of these equations produce conditions on the B-field
parameters $b_a$, due to the fact that $K^K$ only depends on
$v^a={\text{Im}}t^a$, ensuring the reality of the second term,
\begin{equation}
{\text{Im}} \partial_{t^a} W^K = \kappa_{abc}v_b (m_c -m_0 b_c)=0.
\end{equation}
Therefore, $b_c$ is stabilized at $b_c=\frac{m_c}{m_0}$ and vanishes
when $\hat{F}^{\text{bg}}_{(2)}=0$, as claimed above. Of course,
this assumes $m_0 \neq 0$ and also non-vanishing $v_b$ and
$\kappa_{abc}$. This leads us to the real part of equations
(\ref{eq:Kahlereq}). We will show that these yield $h_-^{(1,1)}$
equations to determine the $h_-^{(1,1)}$ moduli fields $v^a$ or
equivalently the $\gamma^a$ used in the discussion earlier. They
read
\begin{equation}
{\text{Re}}(\partial_{t^a} W^K) +
{\text{Im}}(\partial_{t^a} K^K){\text{Im}}(W^K)=0.
\end{equation}
More explicitly, we have
\begin{equation}
\label{eq:explicit} (4 e_a m_0 + 2 \kappa_{apq} m^p m^q + 3 m_0^2
\kappa_{apq} v^p v^q) \kappa_{def}v^d v^e v^f + (6 m_0 e_d v^d + 3
\kappa_{def} v^d m^e m^f) \kappa_{apq}v^p v^q =0,
\end{equation}
where we made frequent use of the equations for the $b_a$ parameters
(see above). Multiplying by $v^a$ and summing over $a$ leads
to\footnote{Solving equation (\ref{eq:explicit}) directly gives no
solution with $vol_{(6)} \neq 0$ and any of the $v_a=0$.}
\begin{equation}
10 m_0 e_d v^d + 5 \kappa_{def} v^d m^e m^f + 3 m_0^2 \kappa_{def}
v^d v^e v^f =0.
\end{equation}
This gives us one quadratic equation for every $v_a$, thus
generically fixing all the K{\"a}hler structure moduli, namely
\begin{equation}
\label{eq:kaehler} 10 m_0 e_a + 5 \kappa_{abc} m^b m^c + 3 m_0^2
\kappa_{abc}
 v^b v^c =0.
\end{equation}

\subsection{Application to the $T^6/{\mathbb Z}_4$ model}
We start out by neglecting the twisted sector to show that we can
reproduce the results found above. Then we discuss the details of
the twisted sector and derive the results for all moduli.\\
\noindent {\bf Complex structure equations.} Combining equations
(\ref{eq:cstructure}) and (\ref{eq:Omega}) we get\footnote{Recall
that we have normalized $\Omega$ s.t. $\ic \int \Omega \wedge
\bar{\Omega} =1$ so that $K^{\text{cs}}=0$.}
\begin{equation}
{\label{eq:U2}} -\frac{4 p_0}{\sqrt{U_2}} = - 8 \sqrt{U_2} p_1=:Q_0.
\end{equation}
Assuming that we satisfy the tadpole cancelation conditions
$p_0=p_1\equiv p$ implies that the complex structure is fixed at
$U_2=\frac{1}{2}$. Since $Q_0=- 4 \sqrt{2} p$, the dilaton
(cf.~(\ref{eq:dilaton})) gets fixed at
\begin{equation}
e^{-\hat{\phi}}= -\frac{\sqrt{2}}{5} \frac{m_0}{p} \sqrt{vol_{(6)}}.
\end{equation}
Note that this implies that $\sign[m_0 p]=-1$. \\
The axions as derived above in (\ref{eq:axion4d}) satisfy
\begin{equation}
p_0 \xi^0 + p_1 \xi^1 = e_0 + e_a b_a + \frac{1}{2} \kappa_{abc} m_a
(b_b b_c - v_b v_c) - \frac{m_0}{6} \kappa_{abc} (b_a b_b b_c - 3
b_a v_b v_c),
\end{equation}
which agrees with (\ref{eq:axion}) for $b_a=\frac{m_a}{m_0}$. \\
\noindent {\bf K{\"a}hler structure equations.} The equations
(\ref{eq:kaehler}) yield the following result for the untwisted
K{\"a}hler moduli,
\begin{align*}
v_1 &= \pm \sqrt{\frac{10}{3}} \frac{\hat{e}_2
\sqrt{\hat{e}_3}}{\sqrt{\kappa m_0} \sqrt{-2 \hat{e}_1 \hat{e}_2 +
\hat{e}_4^2}},\\
v_2 &= \pm \sqrt{\frac{10}{3}} \frac{\hat{e}_1 \sqrt{\hat{e}_3}}
{\sqrt{\kappa m_0} \sqrt{-2 \hat{e}_1 \hat{e}_2 +\hat{e}_4^2}},\\
v_3 &= \mp \sqrt{\frac{5}{6}} \frac{\sqrt{-2 \hat{e}_1 \hat{e}_2+
\hat{e}_4^2}}{\sqrt{\kappa m_0} \sqrt{\hat{e}_3}},\\
v_4 &= \mp \sqrt{\frac{10}{3}} \frac{\hat{e}_4 \sqrt{\hat{e}_3}}{
\sqrt{\kappa m_0} \sqrt{-2 \hat{e}_1 \hat{e}_2 + \hat{e}_4^2}},\\
\end{align*}
where we have defined shifted fluxes invariant under the shifts of
$t_a$\footnote{Remember that there is a modular transformation that
shifts the axions $b_a$ by one.}
\begin{equation}
\hat{e}_i \equiv e_i + \frac{\kappa_{ijk} m_j m_k}{2 m_0}.
\end{equation}
For this solution to have a geometrical interpretation, we have to
demand that\\ $\sign{[m_0 (-2 \hat{e}_1 \hat{e}_2 +
\hat{e}_4^2)]}=\sign{[\hat{e}_3]}$, $v_3>0$ and $(2 \hat{e}_1
\hat{e}_2 - \hat{e}_4^2)>0$. Comparing this with the solution found
in (\ref{eq:solution1}) we see that the additional constraint
$\sign{[m_0 e_3]}<0$ is required for this solution to be supersymmetric.\\
To look at one explicit supersymmetric large volume and small string
coupling example, we use the flux quantization condition
(\ref{eq:quant}) to express the results in terms of flux integers.
Taking the limit $f_1=f_2=f_3=f_4=:f \gg 1$ leads to $v_1 = v_2 = 2
v_3 = -v_4 \sim \frac{72}{\kappa^{1/3}} \frac{\alpha'}
{\sqrt{|f_0|}} \sqrt{f}$. Therefore, for the internal volume, the
string coupling and the potential we get
\begin{align}
vol_{(6)}&= \kappa v_3 (v_1v_2 -\frac{1}{2}{v_4}^2)\sim 9 \times
10^4 \frac{(\alpha')^3}{|f_0|^{3/2}} \: f^{3/2},\\
g_s &= e^{\hat{\phi}} \sim 4 \left| \frac{h}{f_0^{1/4}} \right| \: f^{-3/4},\\
V_{\text{eff}} &= -\sqrt{\frac{3}{10}} \, \frac{243}{1600 \pi^{8}}
\sqrt{\left| \frac{f_0^5}{f^9} \right|} \frac{h^4}{(\alpha')^4} \sim
-9 \times 10^{-6} \sqrt{\left| f_0^5 \right|}
\frac{h^4}{(\alpha')^4} \: f^{-9/2}.
\end{align}
\noindent {\bf Gauge redundancies and counting of solutions.}
An interesting question is to ask how many physically different solutions
there are for different values of the K{\"a}hler axions $b_a=\frac{m_a}{m_0}$.
There are certain modular transformations of infinite order that act as shifts
on the axions and relate equivalent vacua \cite{DeWolfe:2005uu}.
A integer shift of the K{\"a}hler axions
\begin{equation}
b_a \rightarrow b_a + u_a, \quad u_a \in {\mathbb Z}, \forall a,
\end{equation}
corresponds to a shift of the $\hat{F}_2$ flux $m_a \rightarrow m_a
+ u_a m_0$. Now, since $|m_0|$ is (almost) fixed by tadpole
cancelation, we see that physically inequivalent choices for $m_a$
(and thus $b_a$) are defined modulo $|m_0|$. Consequently, once
$m_0$ is fixed there are at most two different
inequivalent solutions for different values of the $b_a$.\\
\noindent {\bf Validity of approximations.}
In order for the low energy supergravity approximation (leading order in $\alpha'$)
 to be valid we have to make sure that the dimensionless expansion parameter
\begin{equation}
\frac{\alpha'}{R^2}\sim f^{-1/2}\ll 1.
\end{equation}
Moreover, we also want the string coupling to be small enough to be in a
perturbative regime where we can safely neglect quantum (string loop)
corrections. As we have observed above, $g_s \sim f^{-3/4}$. Therefore, by
choosing $f \gg 1$ sufficiently large, we can ensure both conditions
simultaneously.\\
Another important issue is the backreaction of the fluxes on the
geometry: Namely, in the presence of background fluxes, the internal
space is strictly speaking no longer a Calabi-Yau orientifold.
However, we want to make sure that the low energy spectrum we
assumed is still correct. For this to be true we must check that the
mass scale of the (canonically normalized) K{\"a}hler moduli is
sufficiently small compared to the mass scale of the massive
Kaluza-Klein modes ($m_{\text{KK}}\sim \frac{1}{R}$) which we
neglected. Performing the calculations in the 4D Einstein frame, we
find
\begin{equation}
m_{{\tilde v}_a} \sim f^{-9/4} \ll m_{\text{KK}}\sim f^{-1/4},
\end{equation}
where ${\tilde v}_a := \frac{\delta v_a}{\kappa_{10}<v_a>}$ is
normalized to give a canonical kinetic term in the Lagrangian.
Clearly, their masses will be much smaller than the Kaluza-Klein
masses if we choose $f \gg 1$ large.

\noindent
\section{Moduli stabilization in the twisted sectors}\label{sec:twisted}
\noindent {\bf Fixed point structure and exceptional divisors.}
After having described the moduli stabilization in the untwisted
sector, it remains to investigate the stabilization of the blow-up
modes in the twisted sectors. Therefore let us briefly summarize the
fixed point structure of our orientifold model (table 6). \vskip
0.8cm \vbox{ \centerline{\vbox{ \hbox{\vbox{\offinterlineskip
\def\tablespace{height2pt&\omit&&\omit&&\omit
&&\omit&\cr}
\def\tablerule{\tablespace\noalign{\hrule}\tablespace}
\def\tableruleA{\tablespace\noalign{\hrule height1pt}\tablespace}
\hrule\halign{&\vrule#&\strut\hskip0.2cm\hfill #\hfill\hskip0.2cm\cr
&sector:  && untwisted && $\Theta,\Theta^3$-twisted &&
$\Theta^2$-twisted && $\sum$ &\cr \tablerule & fixed points/type: &&
--- && 16 ${\mathbb Z}_4$ && 12 ${\mathbb Z}_2$+ 4 ${\mathbb Z}_4$
(${\mathbb Z}_2$) && --- &\cr \tablerule & complex structure:
     && 1 && --- && 6+0 && 1+6 &\cr
\tablerule & K{\"a}hler:  && $5 \rightarrow 4 $(odd)&& $16
\rightarrow 12$ && $6+4 \rightarrow 5+4$ && $5+26 \rightarrow 4 +
21$ &\cr }\hrule}}}} \centerline{ \hbox{{\bf Table Table 6:}{\it ~~
List of moduli before and after orientifold projection.}}} } \vskip
0.5cm \noindent The exceptional divisor can be determined as
follows: We start by modding out the $T^6 = T^2_{(1)} \times
T^2_{(2)} \times T^2_{(3)}$ by the ${\mathbb Z}_2$-action
$\Theta^2$. This yields 16 singularities of type ${\mathbb
C}^2/{\mathbb Z}_2 \times T^2_{(3)}$, whose blow-up is given by 16
${\mathbb CP}^1 \times T^2_{(3)}$. In a second step, we mod out this
blown-up space $\widetilde{T^6/{\mathbb Z}_2}$ by the ${\mathbb
Z}_2$-action $\Theta$. The ${\mathbb CP}^1$s located at ${\mathbb
Z}_2$ fixed points of the first two tori (cf. figure 2) get mapped
into each other by $\Theta$. Moreover, the two ${\mathbb Z}_2$ fixed
points of the second torus are identified under the orientifold
involution $\sigma$. This leaves us with $6 \rightarrow 5 \;
{\mathbb CP}^1 \times T^2_{(3)}$ that contribute to the twisted
K{\"a}hler moduli. Furthermore, the 6 ${\mathbb CP}^1$s at the
${\mathbb Z}_2$ fixed points can be tensored with the two 1-cycles
on the third torus to yield 12 twisted 3-cycles of topology $S^2
\times S^1$ (which contribute 6 twisted complex structure moduli).
The 4 ${\mathbb CP}^1$s sitting at the ${\mathbb Z}_4$ fixed points
of the first two tori remain invariant under this action and
contribute 4 K{\"a}hler moduli (the sizes of the ${\mathbb CP}^1$s)
to the twisted sectors. The 16 fixed loci of the $\Theta$-action are
${\mathbb CP}^1 \times \{ \text{point} \}$, where $\{
\text{point}\}$ denotes one of the fixed points of the third torus
(cf. figure 2). Two of these get identified by $\sigma$. Blowing-up
results in $16 \rightarrow 12 \; {\mathbb CP}^1 \times {\mathbb
CP}^1$, which give
us the 12 K{\"a}hler moduli from the $\Theta^1,\Theta^3$ sectors. \\
\noindent {\bf Intersection numbers.} In order to solve the F-term
conditions for the twisted K{\"a}hler moduli, we need to calculate
the various triple intersection numbers of the blow-up cycles. The
results are listed in table 7 below. \vskip 0.8cm \vbox{
\centerline{\vbox{ \hbox{\vbox{\offinterlineskip
\def\tablespace{height2pt&\omit&&\omit&&\omit&\cr}
\def\tablerule{\tablespace\noalign{\hrule}\tablespace}
\def\tableruleA{\tablespace\noalign{\hrule height1pt}\tablespace}
\hrule\halign{&\vrule#&\strut\hskip0.2cm\hfill #\hfill\hskip0.2cm\cr
& divisor && intersection type && intersection number &\cr
\tablerule & $T={\mathbb CP}^1 \times T^2_{(3)}$
     && $T \circ T \circ T$  && 0 &\cr
\tablerule & $T={\mathbb CP}^1 \times T^2_{(3)}$ && $T \circ T \circ
\left[U=T^2_{(1)}\times T^2_{(2)}\right]$ && $\beta=-2$ &\cr
\tablerule & $T'={\mathbb CP}^1 \times {\mathbb CP}^1$ && $T' \circ
T' \circ T'$ && $\alpha=8$ &\cr }\hrule}}}} \centerline{ \hbox{{\bf
Table 7:}{\it ~~ List of intersection numbers.}}} } \vskip 0.5cm
\noindent These results can be used to extend the F-term equations
discussed above to include the twisted moduli.\\
It is important to note that there must be a hierarchy between the
untwisted and twisted K{\"a}hler moduli,
\begin{equation}
|m_0| \ll |e_A| \ll |e_a|,
\end{equation}
in order to remain within the K{\"a}hler cone \cite{DeWolfe:2005uu}.
This is the reason why, although there are non-vanishing
intersection numbers linking the twisted sectors to the untwisted
sector, the values at which the untwisted K{\"a}hler moduli are
stabilized will not significantly change compared to the analysis of
only the untwisted sector above.\\
\noindent {\bf Solutions to K{\"a}hler structure equations.} For the
$b_a$ we have the same solutions as above $b_a=\frac{m_a}{m_0}$ where
$a$ now runs from 0 to 26.\\
For the $v_a$ we have to solve the equations~(\ref{eq:kaehler}). The
solution is
\begin{align}
v_1 &= \pm \sqrt{\frac{10}{3}} \frac{\hat{e}_2 \sqrt{\hat{e}_3}}
{\sqrt{\kappa m_0}\sqrt{(-2 \hat{e}_1 \hat{e}_2 + {\hat{e}_4}^2) -
\frac{\kappa}{\beta} ({\hat{e}_5}^2 + \ldots +{\hat{e}_{14}}^2)}},\\
\nonumber v_2 &= \pm \sqrt{\frac{10}{3}} \frac{\hat{e}_1
\sqrt{\hat{e}_3}} {\sqrt{\kappa m_0} \sqrt{(-2 \hat{e}_1 \hat{e}_2 +
{\hat{e}_4}^2) - \frac{\kappa}{\beta} ({\hat{e}_5}^2 + \ldots
+{\hat{e}_{14}}^2)}},\\\nonumber v_3 &= \mp \sqrt{\frac{5}{6}}
\frac{\sqrt{(-2 \hat{e}_1 \hat{e}_2 + {\hat{e}_4}^2) -
\frac{\kappa}{\beta} ({\hat{e}_5}^2 + \ldots +{\hat{e}_{14}}^2)}}
{\sqrt{\kappa m_0} \sqrt{\hat{e}_3}},\\\nonumber v_4 &= \mp
\sqrt{\frac{10}{3}} \frac{\hat{e}_4 \sqrt{\hat{e}_3}} {\sqrt{\kappa
m_0}\sqrt{(-2 \hat{e}_1 \hat{e}_2 + {\hat{e}_4}^2) -
\frac{\kappa}{\beta} ({\hat{e}_5}^2 + \ldots +{\hat{e}_{14}}^2)}},
\\\nonumber v_5 &= \pm \sqrt{\frac{10}{3}} \frac{\hat{e}_5
\sqrt{\kappa \hat{e}_3}}{\sqrt{m_0} \, \beta \, \sqrt{(-2 \hat{e}_1
\hat{e}_2 +{\hat{e}_4}^2)-\frac{\kappa}{\beta} ({\hat{e}_5}^2 +
\ldots +{\hat{e}_{14}}^2)}},\\\nonumber & \qquad \vdots\\\nonumber
v_{14} &= \pm \sqrt{\frac{10}{3}} \frac{\hat{e}_{14} \sqrt{\kappa
\hat{e}_3}} {\sqrt{m_0} \, \beta \, \sqrt{(-2 \hat{e}_1 \hat{e}_2
+{\hat{e}_4}^2) - \frac{\kappa}{\beta} ({\hat{e}_5}^2 + \ldots
+{\hat{e}_{14}}^2)}},\\\nonumber v_{15} &= \pm \sqrt{\frac{10}{3}}
\sqrt{-\frac{\hat{e}_{15}}{\alpha m_0}},\\\nonumber & \qquad
\vdots\\\nonumber v_{26} &= \pm \sqrt{\frac{10}{3}}
\sqrt{-\frac{\hat{e}_{26}}{\alpha m_0}}.
\end{align}
As before, there are some additional conditions on the relative
signs of the fluxes. To ensure reality of the K{\"a}hler moduli, we
need to have
\begin{align}
&\sign{[\left(\frac{\hat{e}_3}{m_0 ((-2 \hat{e}_1 \hat{e}_2 +
{\hat{e}_4}^2)- \frac{\kappa}{\beta} ({\hat{e}_5}^2 + \ldots +
{\hat{e}_{14}}^2))}\right)]}>0,\\
&\sign{[\left(\frac{\hat{e}_A}{\alpha m_0}\right)]}<0, \quad \forall
A=15,\ldots,26.
\end{align}
The volume and the string coupling constant are
\begin{align}
vol_{(6)}&= \frac{1}{6} \kappa_{abc} v_a v_b v_c\\
&= v_3 (\kappa v_1 v_2 -\frac{\kappa}{2}{v_4}^2 + \frac{\beta}{2}
\sum_{A=5}^{14} v_A^2) + \frac{\alpha}{6}
\sum_{A=15}^{26} v_A^3\\
&= \frac{5}{3} \sqrt{\frac{5}{6}} \sqrt{\left| \frac{\hat{e}_3 ((2 \hat{e}_1 \hat{e}_2
-\hat{e}_4^2)+\frac{\kappa}{\beta} (\hat{e}_5^2 + \ldots +\hat{e}_{14}^2))}{\kappa
m_0^3} \right|}+ \frac{\alpha}{6} \sum_{A=15}^{26} \left( - \frac{10
\hat{e}_A}{3
\alpha m_0} \right)^{3/2},\\
g_s &= e^{\hat{\phi}} =-\frac{5}{\sqrt{2}} \frac{p}{m_0}
\frac{1}{\sqrt{vol_{(6)}}}.
\end{align}
Due to the hierachy of fluxes mentioned above, the results for the
untwisted sector do not deviate substantially from those obtained
without taking the twisted sector into account.\\
\noindent {\bf Twisted complex structure moduli.} As we saw above,
including the twisted sector we now have 7 complex structure moduli
to stabilize. The holomorphic 3-form is $\Omega (\tilde{z}) =
Z^{K}(\tilde{z}) a_{K} - {\mathcal F}_{K}(\tilde{z})b^{K},
{K}=0,\ldots,6$. Equation ({\ref{eq:U2}}) is still valid if we fix
the normalization of $\Omega$ such that $\ic \int \Omega \wedge
{\bar{\Omega}}=1$. For the twisted complex structure the $p_{K},
{K}=2,\ldots,6$ are not constrained by the tadpole conditions. We
can for example choose them to be $p_{K}=0, {K}=2,\ldots,6$ which
would fix the corresponding complex structures ${\text{Im}({\mathcal
F}_{K})}=0$. If we choose any of the $p_{K}, {K}=2,\ldots,6$ to be
non zero, the corresponding complex structure is fixed as
\begin{equation}
{\text{Im}({\mathcal F}_{K})}= -\frac{p_{K}}{4 \sqrt{2} \, p}.
\end{equation}
The axions as derived above in (\ref{eq:axion4d}) satisfy
\begin{equation}
\sum_{h^{(2,1)}=0}^7 p_i \xi^i = e_0 + \frac{e_a m_a}{m_0} +
\frac{\kappa_{abc} m_a m_b m_c}{3 m_0^2},
\end{equation}
where we have used $b_a=\frac{m_a}{m_0}$ and $a,b,c$ run from 1 to
26.

\section{Conclusions and Outlook}\label{sec:conclusions}

In this note we have worked out the moduli stabilization for a
specific type IIA orientifold model, namely an orientifolded
$T^6/{\mathbb Z}_4$ orbifold. The hope is that it will now be
possible to add certain ingredients (D6-branes) in order to build a
(semi-)realistic model which combines an MSSM(-like) particle
content with realistic cosmological features, e.g., $\Lambda >0$,
without introducing new, unfixed moduli. This will be addressed in a
forthcoming paper \cite{Ihl:2006xx}. A summary of what needs to be
done is outlined in the following. We would like to lift the stable
AdS vacua derived above to meta-stable dS vacua in a controlled way.
There has been a renewed interest in recent literature in
investigating the possibility of D-term induced spontaneous
supersymmetry breaking
\cite{Villadoro:2006ia,Villadoro:2005yq,Conlon:2006tq,Martucci:2006ij}.
In analogy to the type IIB case, where U(1) gauge field fluxes on
$D7$-branes (magnetized $D7$-branes) wrapping 4-cycles in the
internal space were proposed as a means to generate D-terms (and
F-terms) which spontaneously break $N=1$ supersymmetry
\cite{Burgess:2003ic, Jockers:2005zy}, we propose to use gauge field
fluxes on $D6$-branes to induce similar terms in the type IIA setup
\cite{Ihl:quali,Villadoro:2006ia}. However, no concrete, viable
stringy realization of a D-term uplift to a meta-stable dS vacuum
has been found so far. According to \cite{Villadoro:2005yq}, a
necessary prerequisite for constructing D-term contributions fully
consistent with supergravity constraints is the existence of unfixed
axions that can participate in a supersymmetric Higgs mechanism
(St\"uckelberg mechanism) to form a massive U(1) vector. As we have
seen above, such unfixed (complex structure) axions exist in our
model. Therefore, it would be interesting to see if we can
consistently apply D-term supersymmetry
breaking in this class of models.\\
Moreover, we would like to incorporate stacks of intersecting
$D6$-branes \cite{Marchesano:2006ns} so as to build a
(semi-)realistic particle spectrum featuring standard model or MSSM
(-like) gauge groups. It was demonstrated in
\cite{Blumenhagen:2002gw} that the $T^6/{\mathbb Z}_4$ orientifold
model under consideration can give rise to interesting particle
phenomenology, such as a 3-generation Pati-Salam model, utilizing
supersymmetric configurations of fractional $D6$-branes. However,
since we are working in the framework of massive type IIA theory,
the presence of $D8$-branes renders $D6$-brane configurations that
preserve some supersymmetry much less generic
(cf.~\cite{Behrndt:2003ih}). Therefore, a careful investigation of
all the constraints is crucial to fully understand the
phenomenological viability of such models. This interesting topic
will be the subject of future study.

\section*{Acknowledgements}
It is a pleasure to thank Aaron Bergman, Jacques Distler, Raphael Flauger,
Sonia Paban and Uday Varadarajan for many helpful comments, discussions and
encouragement.
The research of the authors is based upon work supported by the National
Science Foundation under Grant Nos. PHY-0071512 and PHY-0455649,
and with grant support from the US Navy, Office of Naval Research,
Grant Nos. N00014-03-1-0639 and N00014-04-1-0336, Quantum Optics
Initiative.

\bibliography{finalv4}

\providecommand{\href}[2]{#2}\begingroup\raggedright\begin{thebibliography}{10}
\bibitem{Seljak:2004xh}
U.~Seljak {\it et al.},
``Cosmological parameter analysis including SDSS Ly-alpha forest and
galaxy bias: Constraints on the primordial spectrum of fluctuations,
neutrino mass, and dark energy,''
arXiv:astro-ph/0407372.

\bibitem{Riess:2004nr}
A.~G.~Riess {\it et al.}  [Supernova Search Team Collaboration],
``Type Ia Supernova Discoveries at $z>1$ From the Hubble Space Telescope:
Evidence for Past Deceleration and Constraints on Dark Energy
Evolution,''
Astrophys.\ J.\  {\bf 607}, 665 (2004)
[arXiv:astro-ph/0402512].

\bibitem{Spergel:2003cb}
D.~N.~Spergel {\it et al.}  [WMAP Collaboration],
``First Year Wilkinson Microwave Anisotropy Probe (WMAP) Observations:
Determination of Cosmological Parameters,''
Astrophys.\ J.\ Suppl.\  {\bf 148}, 175 (2003)
[arXiv:astro-ph/0302209].

\bibitem{Fischler:2001yj}
W.~Fischler, A.~Kashani-Poor, R.~McNees and S.~Paban,
``The acceleration of the universe, a challenge for string theory,''
JHEP {\bf 0107}, 003 (2001)
[arXiv:hep-th/0104181].

\bibitem{Hellerman:2001yi}
S.~Hellerman, N.~Kaloper and L.~Susskind,
``String theory and quintessence,''
JHEP {\bf 0106}, 003 (2001)
[arXiv:hep-th/0104180].

\bibitem{Kachru:2003aw}
S.~Kachru, R.~Kallosh, A.~Linde and S.~P.~Trivedi,
``De Sitter vacua in string theory,''
Phys.\ Rev.\ D {\bf 68}, 046005 (2003)
[arXiv:hep-th/0301240].

\bibitem{Giddings:2001yu}
S.~B.~Giddings, S.~Kachru and J.~Polchinski,
``Hierarchies from fluxes in string compactifications,''
Phys.\ Rev.\ D {\bf 66}, 106006 (2002)
[arXiv:hep-th/0105097].

\bibitem{Gukov:1999ya}
S.~Gukov, C.~Vafa and E.~Witten,
``CFT's from Calabi-Yau four-folds,''
Nucl.\ Phys.\ B {\bf 584}, 69 (2000)
[Erratum-ibid.\ B {\bf 608}, 477 (2001)]
[arXiv:hep-th/9906070].

\bibitem{Witten:1996bn}
E.~Witten,
``Non-Perturbative Superpotentials In String Theory,''
Nucl.\ Phys.\ B {\bf 474}, 343 (1996)
[arXiv:hep-th/9604030].

\bibitem{Veneziano:1982ah}
G.~Veneziano and S.~Yankielowicz,
``An Effective Lagrangian For The Pure N=1 Supersymmetric Yang-Mills
Theory,''
Phys.\ Lett.\ B {\bf 113}, 231 (1982).

\bibitem{Taylor:1982bp}
T.~R.~Taylor, G.~Veneziano and S.~Yankielowicz,
``Supersymmetric QCD And Its Massless Limit: An Effective Lagrangian
Analysis,''
Nucl.\ Phys.\ B {\bf 218}, 493 (1983).

\bibitem{Berglund:2005dm}
P.~Berglund and P.~Mayr,
``Non-perturbative superpotentials in F-theory and string duality,''
arXiv:hep-th/0504058.

\bibitem{Douglas:2003um}
M.~R.~Douglas,
``The statistics of string / M theory vacua,''
JHEP {\bf 0305}, 046 (2003)
[arXiv:hep-th/0303194].

\bibitem{Denef:2005mm}
F.~Denef, M.~R.~Douglas, B.~Florea, A.~Grassi and S.~Kachru,
``Fixing all moduli in a simple F-theory compactification,''
arXiv:hep-th/0503124.

\bibitem{Gorlich:2004qm}
L.~G{\"o}rlich, S.~Kachru, P.~K.~Tripathy and S.~P.~Trivedi,
``Gaugino condensation and nonperturbative superpotentials in flux
compactifications,''
arXiv:hep-th/0407130.

\bibitem{Lust:2005dy}
D.~Lust, S.~Reffert, W.~Schulgin and S.~Stieberger,
``Moduli stabilization in type IIB orientifolds. I: Orbifold limits,''
arXiv:hep-th/0506090.

\bibitem{Reffert:2005mn}
S.~Reffert and E.~Scheidegger,
``Moduli stabilization in toroidal type IIB orientifolds,''
Fortsch.\ Phys.\  {\bf 54}, 462 (2006)
[arXiv:hep-th/0512287].

\bibitem{Balasubramanian:2004uy}
V.~Balasubramanian and P.~Berglund,
``Stringy corrections to Kaehler potentials, SUSY breaking, and the
cosmological constant problem,''
JHEP {\bf 0411}, 085 (2004)
[arXiv:hep-th/0408054].

\bibitem{Balasubramanian:2005zx}
V.~Balasubramanian, P.~Berglund, J.~P.~Conlon and F.~Quevedo,
``Systematics of moduli stabilisation in Calabi-Yau flux compactifications,''
JHEP {\bf 0503}, 007 (2005)
[arXiv:hep-th/0502058].

\bibitem{Conlon:2005ki}
J.~P.~Conlon, F.~Quevedo and K.~Suruliz,
``Large-volume flux compactifications: Moduli spectrum and D3/D7 soft
supersymmetry breaking,''
arXiv:hep-th/0505076.

\bibitem{Villadoro:2005cu}
G.~Villadoro and F.~Zwirner,
``N = 1 effective potential from dual type-IIA D6/O6 orientifolds with
general fluxes,''
JHEP {\bf 0506}, 047 (2005)
[arXiv:hep-th/0503169].

\bibitem{DeWolfe:2005uu}
O.~DeWolfe, A.~Giryavets, S.~Kachru and W.~Taylor,
``Type IIA moduli stabilization,''
arXiv:hep-th/0505160.

\bibitem{Kachru:2004jr}
S.~Kachru and A.~K.~Kashani-Poor,
``Moduli potentials in type IIA compactifications with RR and NS flux,''
JHEP {\bf 0503}, 066 (2005)
[arXiv:hep-th/0411279].

\bibitem{Camara:2005dc}
P.~G.~C{\'a}mara, A.~Font and L.~E.~Ib{\'a}{\~n}ez,
``Fluxes, moduli fixing and MSSM-like vacua in a simple IIA orientifold,''
arXiv:hep-th/0506066.

\bibitem{Saueressig:2005es}
F.~Saueressig, U.~Theis and S.~Vandoren,
``On de Sitter Vacua in Type IIA Orientifold Compactifications,''
arXiv:hep-th/0506181.

\bibitem{Dixon:1985jw}
L.~J.~Dixon, J.~A.~Harvey, C.~Vafa and E.~Witten,
``Strings On Orbifolds,''
Nucl.\ Phys.\ B {\bf 261}, 678 (1985).

\bibitem{Dixon:1986jc}
L.~J.~Dixon, J.~A.~Harvey, C.~Vafa and E.~Witten,
``Strings On Orbifolds. 2,''
Nucl.\ Phys.\ B {\bf 274}, 285 (1986).

\bibitem{Blumenhagen:1999ev}
R.~Blumenhagen, L.~G{\"o}rlich and B.~K{\"o}rs,
``Supersymmetric 4D orientifolds of type IIA with D6-branes at angles,''
JHEP {\bf 0001}, 040 (2000)
[arXiv:hep-th/9912204].

\bibitem{Blumenhagen:2002gw}
R.~Blumenhagen, L.~G{\"o}rlich and T.~Ott,
``Supersymmetric intersecting branes on the type IIA T**6/Z(4) orientifold,''
JHEP {\bf 0301}, 021 (2003)
[arXiv:hep-th/0211059].

\bibitem{Blumenhagen:2002wn}
R.~Blumenhagen, V.~Braun, B.~K{\"o}rs and D.~L{\"u}st,
``Orientifolds of K3 and Calabi-Yau manifolds with intersecting D-branes,''
JHEP {\bf 0207}, 026 (2002)
[arXiv:hep-th/0206038].

\bibitem{Blumenhagen:2004di}
R.~Blumenhagen, J.~P.~Conlon and K.~Suruliz,
``Type IIA orientifolds on general supersymmetric Z(N) orbifolds,''
JHEP {\bf 0407}, 022 (2004)
arXiv:hep-th/0404254].

\bibitem{Grimm:2004ua}
T.~W.~Grimm and J.~Louis,
``The effective action of type IIA Calabi-Yau orientifolds,''
Nucl.\ Phys.\ B {\bf 718}, 153 (2005)
[arXiv:hep-th/0412277].

\bibitem{Acharya:2002ag}
B.~Acharya, M.~Aganagic, K.~Hori and C.~Vafa,
``Orientifolds, mirror symmetry and superpotentials,''
arXiv:hep-th/0202208.

\bibitem{Grimm:2005fa}
T.~W.~Grimm,
``The effective action of type II Calabi-Yau orientifolds,''
Fortsch.\ Phys.\  {\bf 53}, 1179 (2005)
[arXiv:hep-th/0507153].

\bibitem{Grana:2005jc}
 M.~Gra{\~n}a,
 ``Flux compactifications in string theory: A comprehensive review,''
 arXiv:hep-th/0509003.

\bibitem{Ihl:2006xx}
M.~Ihl, C.~Krishnan, U.~Varadarajan and T.~Wrase, {\sl to appear}.

\bibitem{Burgess:2003ic}
C.~P.~Burgess, R.~Kallosh and F.~Quevedo,
``de Sitter string vacua from supersymmetric D-terms,''
JHEP {\bf 0310}, 056 (2003)
[arXiv:hep-th/0309187].

\bibitem{Jockers:2005zy}
H.~Jockers and J.~Louis,
``D-terms and F-terms from D7-brane fluxes,''
Nucl.\ Phys.\ B {\bf 718}, 203 (2005)
[arXiv:hep-th/0502059].


\bibitem{Ihl:quali}
M.~Ihl,
{\sl Qualifier Talk: Moduli stabilization in type IIA flux compactifications}, 
University of Texas at Austin; November 28, 2005.

\bibitem{Villadoro:2006ia}
G.~Villadoro and F.~Zwirner,
``D terms from D-branes, gauge invariance and moduli stabilization in flux
compactifications,''
arXiv:hep-th/0602120.

\bibitem{Villadoro:2005yq}
G.~Villadoro and F.~Zwirner,
``de Sitter vacua via consistent D-terms,''
Phys.\ Rev.\ Lett.\  {\bf 95}, 231602 (2005)
[arXiv:hep-th/0508167].

\bibitem{Conlon:2006tq}
J.~P.~Conlon,
``The QCD axion and moduli stabilisation,''
arXiv:hep-th/0602233.

\bibitem{Martucci:2006ij}
L.~Martucci,
``D-branes on general N = 1 backgrounds: Superpotentials and D-terms,''
arXiv:hep-th/0602129.

\bibitem{Marchesano:2006ns}
F.~Marchesano,
``D6-branes and torsion,''
arXiv:hep-th/0603210.

\bibitem{Behrndt:2003ih}
K.~Behrndt and M.~Cvetic,
``Supersymmetric intersecting D6-branes and fluxes in massive type IIA
string theory,''
Nucl.\ Phys.\ B {\bf 676}, 149 (2004)
[arXiv:hep-th/0308045].

\end{thebibliography}\endgroup
\bibliographystyle{utphys}
\end{document}